# Limiting energy dissipation induces glassy kinetics in single cell high precision responses

Jayajit Das

Battelle Center for Mathematical Medicine, The Research Institute at the Nationwide Children's Hospital, and, the Biophysics Program, and the Departments of Pediatrics, and, Physics, the Ohio State University, Columbus, OH 43205

Single cells often generate precise responses by involving dissipative out-of-thermodynamic equilibrium processes in signaling networks. The available free energy to fuel these processes could become limited depending on the metabolic state of an individual cell. How does limiting dissipation affect the kinetics of high precision responses in single cells? I address this question in the context of a kinetic proofreading scheme used in a simple model of early time T cell signaling. I show using exact analytical calculations and numerical simulations that limiting dissipation qualitatively changes the kinetics in single cells marked by emergence of slow kinetics, large cell-to-cell variations of copy numbers, temporally correlated stochastic events (dynamic facilitation), and, ergodicity breaking. Thus, constraints in energy dissipation, in addition to negatively affecting ligand discrimination in T cells, can create a fundamental difficulty in interpreting single cell kinetics from cell population level results.

**Introduction**

Living systems are capable of generating surprisingly precise responses in noisy environments(1-5). For example, T cells, a major orchestrator of adaptive immunity in jawed vertebrates, can identify antigen presenting cells (APCs) displaying few pathogenic ligands (1-10 molecules) in the background of tens of thousands of self-ligands (2, 6). High precision responses such as the above are often produced by (free) energy dissipating non-equilibrium thermodynamic processes (2, 3, 5, 7, 8). Therefore, a continuous supply of energy is required to support execution of these processes.

However, the availability of energy (e.g., the ATP pool) for generating high precision cell responses can depend on a number of factors such as, nutrient availability in the local environment(9), activation of specific signaling pathways regulating ATP generation(10), or, prioritization of other cell functions over high precision response (11, 12). For example, activation of the kinase AMPK in T cells during early time signaling events leads to ATP generation that provides energy to execute the signaling processes(10). In another case, similar to the Warburg effect in tumor cells(11), effector T cells employ a less efficient ATP producing glucose metabolism to prioritize cell proliferation(12).

The mechanistic details regarding how variation of energy supply affects energy consuming responses in single cells are not well understood. Relation between energy dissipation, and, speed and error in high precision responses such as kinetic proof reading (KPR) (7, 8, 13)or chemotaxis(4) has been investigated in mathematical models for cell populations. These studies suffer from two major drawbacks: (1) It is unclear to what extent the results obtained at the level of cell populations will generalize to single cells where the signaling kinetics(1) as well as the dissipated energy(14) are affected by intrinsic and extrinsic noise fluctuations(1). (2) Some of these studies are carried out at the steady states of the kinetics. Since energy restriction could slow down the kinetics, the steady states could occur in time scales that are unrealistically large for a biological function of interest. Thus, the steady state results are unlikely to hold in such situations. I address the above issues here in the context of a KPR model of ligand discrimination in single immune cells (T cells) and demonstrate that the single cell responses in the dissipation limited case can be fundamentally different than their counterparts with unrestricted dissipation. The KPR model in the presence of intrinsic and extrinsic noise fluctuations pertaining to single cells has been analyzed previously(15), however, the role of limiting dissipation in affecting the signaling kinetics has not been investigated before.

This concept of KPR, originally proposed by Hopfield(3) and Ninio(5), was applied by McKeithan(16) to explain the remarkable ability of immune cells (such as T and B cells) to sensitively discriminate between closely related antigens. These cells are able to distinguish between similar antigens whose half-lives differ only by few seconds(2, 17). A key biochemical step in McKeithan's scheme is that upon ligand (antigen) unbinding from the receptor any activated state of the receptor is reset to the neutral state (Fig. 1A). While this step increases the sensitivity of the response it also breaks the detailed balance condition(1, 18-20) creating a need for constant (free) energy supply that is dissipated by the system to sustain a non-vanishing probability current in the biochemical network. Biophysical models(2, 15, 21) for antigen discrimination in T cells that provide better agreement with experimental data than the original model proposed by McKeithan use KPR as the core concept and require breakdown of the detailed balance condition and continuous energy dissipation. Therefore, the results obtained here regarding the role of energy dissipation in regulating the kinetics will have implications for these models.

Here, I investigated the role of limiting energy dissipation in a minimal model involving a KPR scheme for ligand discrimination in single T cells using semi-analytical calculations and continuous time Monte Carlo (MC) simulations. Dissipation in single cells is quantified by calculating the rate of entropy production in single microscopic trajectories(22). I specifically investigated cases where the energy pool available for dissipation is either fixed or increases with a constant rate. The results showed, in single cells, limiting dissipation is marked by emergence of slow kinetics, large cell-to-cell variations of signaling kinetics, and, arrest of activated or de-activated signaling states for prolonged time intervals; all of which severely disrupt the sensitive discrimination program in immune cells. Furthermore, the emergent kinetics in dissipation limited situations displays dynamic facilitation(23) and ergodicity breaking(24) bearing an interesting similarity to that in facilitated models for glass formers (e.g., supercooled liquids) below the glass transition temperature(23). The presence of the glassy kinetics points to a fundamental disconnect between the signaling kinetics in single cells and the signaling kinetics obtained by averaging over a cell population. In addition, the results reveal a novel mechanism for emergence of glassy kinetics in non-equilibrium systems that is likely to generalize in a large variety of non-equilibrium systems where kinetic constraints in the dynamics are imposed by limiting dissipation.

**Model**

A minimal model, based on the KPR model proposed by McKeithan(16), was developed to describe early time signaling kinetics in a single T cell (Fig. 1A). However, McKeithan's KPR model and the later modifications of the model(2, 15, 21) do not consider restriction of energy limitation in the system. The novelty of the model constructed here is in its ability to study biochemical kinetics in various dissipation limited situations. In the model, plasma membrane bound T cell receptors (TCRs) interact with antigens or peptide-Major Histocompatibility Complex (pMHC) molecules on APCs with an affinity characterized by a binding ($k_{on}$) and an unbinding rate ($k_{off}$). A single TCR (T) binds a pMHC molecule (M) to form a complex, TM, and, TM then transitions to an activated state T*M. The reaction, TM→ T*M, represents kinase mediated phosphorylation of the tyrosine residues in motifs of amino acids (also known as ITAMs) associated with TCRs(2). The activated state can become deactivated (e.g., TM→ T*M) due to the action of phosphatases(2). Both the activation (occurs at a rate $k_p$) and deactivation (occurs at a rate $k_d$) transitions are assumed to be first order reactions where action of kinases and phosphatases are accounted for implicitly. A key step proposed by McKeithan(16), which I will call the kinetic proofreading (KPR) step, leads to complete deactivation of the activated complex T*M (occurs with a rate $k_{off}'$) upon ligand unbinding, i.e, T*M→ T+M. Unless mentioned, I will assume $k_{off}'=k_{off}$. In order to keep the entropy calculations finite, a transition T+M→T*M (rate $k_1$) is assumed to occur at a much larger time scale than any biologically realistic time scale. I will designate the above model sans the KPR step as the non-KPR (NKPR) model hereafter.

McKeithan(16) analyzed a general form of the above KPR model using deterministic mass action kinetics and showed that the KPR step endows the model with a higher discriminatory power for selecting higher affinity pathogen derived peptide ligands from low affinity naturally occurring ligands (self-ligands) in the host. A similar analysis for the above KPR model (Fig. 1A) showed that for a range of parameters, the steady state concentration of T*M varies as $(1/k_{off})^2$ as opposed to $1/k_{off}$ in the NKPR model (web supplement).

*A. Kinetics and dissipation in the minimal model*: The biochemical kinetics of the copy numbers of the molecular species in the model is subject to intrinsic stochastic fluctuations arising due to the thermal noise(1). I will consider the molecules to be well mixed in a small volume (1μm$^2$ (plasma membrane area) × 0.01μm (depth in the cytosol)) in the membrane proximal region, which is a reasonable approximation. The stochastic kinetics of the biochemical reactions is described by the Master equation (Eq. (1)) in terms of the conditional probability $p(i,t|i_0,0)$ (denoted as $p_i(t)$ from now on for brevity) which is the probability for the system to be in the state $i$ at time $t$ given it started at the state $i_0$ at time $t=0$. $p_i(t)$ follows the kinetics below(1, 18, 20):

$$\frac{dp_i(t)}{dt} = \sum_{j(j \neq i)} [w_i^j p_j(t) - w_j^i p_i(t)] \tag{1}$$

where, $w_j^i$ is the rate of the transition $i \rightarrow j$. I will follow a notation scheme where the system always transitions top→bottom, i.e., from the state in the superscript to that in the subscript. In the model, any state $i$ is specified by a pair of integers, $N_{TM}$ and $N_{T*M}$, denoting copy numbers of the species TM and T*M, respectively. The copy numbers of other two species T and M are related to $N_{TM}$ and $N_{T*M}$ via the total numbers of TCRs ($N_{T0}$) and MHCs ($N_{M0}$) in the model, i.e., $N_{T0} = N_T + N_{TM} + N_{T*M}$ and $N_{M0} = N_M + N_{TM} + N_{T*M}$. Since, $N_{T0}$ and $N_{M0}$ do not change in the biochemical reactions, the stochastic kinetics in the model can be represented by a continuous time random walk (CTRW) (25) model where the random walker moves on a two dimensional square lattice with a lattice spacing of unity. A lattice point (n,m) in the model denotes the biochemical state with $N_{TM}$ = n and $N_{T*M}$=m. When a reaction occurs, the random walker instantaneously steps to one of the four nearest neighbor sites from the current site (n,m). The walker waits for a duration $\tau$ at the current site (n,m) before taking the next step where the values of the waiting time $\tau$ are drawn from a continuous probability density function determined by Eq. (1). A particular stochastic trajectory in the CTRW model describes the kinetics of the molecular species in a single cell, and, since I assume the total numbers of TCRs and pMHCs do not change from cell to cell, averaging over an ensemble of stochastic trajectories (denoted by the angular brackets, $\langle \cdots \rangle$, hereafter) also implies averaging over a cell population. The average over the cell population is equivalent to an average over $p(i,t|i_0,0)$ when the cell population contains a large number of single cells. The CTRW representation will be utilized later for analyzing stochastic trajectories from MC simulations.

The energy dissipation is characterized by the entropy production in the kinetics. The system entropy is defined as, $S_{sys} = -\sum_i p_i(t)\ln[p_i(t)]$(18-20, 26), where the sum over i also represents a sum over single cells in a cell population (or an ensemble of stochastic trajectories). $S_{sys}$ follows the kinetics(18-20, 26),

$$\frac{dS_{sys}(t)}{dt} = -\underbrace{\sum_{\substack{i,j \\ j \neq i}} w_i^j p_j \ln\left(\frac{w_j^i p_i}{w_i^j p_j}\right)}_{\frac{dS_{total}}{dt}} - \underbrace{\sum_{\substack{i,j \\ j \neq i}} w_i^j p_j \ln\left(\frac{w_j^i}{w_i^j}\right)}_{\frac{dS_{med}}{dt}} = \frac{dS_{total}}{dt} - \frac{dS_{med}}{dt}$$

(2)

According to Eq. (2), the entropy $S_{total}$ never decreases(19, 26), i.e., $dS_{total}/dt \geq 0$, and thus quantifies dissipation in the system. In the steady state, $dS_{sys}/dt=0$, and, consequently, $dS_{total}/dt=dS_{med}/dt$. $dS_{med}/dt$ denotes the rate of entropy exchange between the system and the reservoir. Dissipative systems (e.g., the minimal model with the KPR step) receive entropy from the reservoir at a fixed rate (i.e., $dS_{med}/dt \approx v > 0$) in the steady state to maintain a constant probability current(19, 20, 26). In contrast, $dS_{med}/dt=0$ at the steady state in the NKPR model due to the vanishing steady state probability current in the absence of the KPR step (web supplement). Thus, the steady state kinetics in the NKPR state is dissipationless, i.e., $dS_{total}/dt=0$. Following Seifert(19), it is possible to construct different entropies for single stochastic trajectories or kinetics in single cells that correspond to, $S_{total}$, $S_{med}$, and, $S_{sys}$, defined above. Due to the relevance of the medium entropy ($S_{med}$) in characterizing dissipation in a cell population I will focus on the entropy exchanges for single cells or single stochastic trajectories. For a sequence of $N$ biochemical reaction events in a time interval t the total amount of medium entropy that flows into the system from the reservoir is given by(19),

$$Q(t) = \sum_{\alpha=1}^{N} \left( \Delta s_m \right)_{i_\alpha}^{j_\alpha} = \sum_{\alpha=1}^{N} \ln \left( \frac{w_{i_\alpha}^{j_\alpha}}{w_{j_\alpha}^{i_\alpha}} \right),$$

(3)

where, the αth stochastic transition, $j_\alpha \rightarrow i_\alpha$, occurring at time $t_\alpha$ is associated with an entropy flow, $\left( \Delta s_m \right)_{i_\alpha}^{j_\alpha} = \ln(w_{i_\alpha}^{j_\alpha} / w_{j_\alpha}^{i_\alpha})$ (19), from the reservoir to the system. $\left( \Delta s_m \right)_i^j$ will be denoted by $\Delta s_i^j$ from now on. $Q(t)$ in Eq. (3) is also a stochastic variable that varies between stochastic trajectories or single cells and will be used to quantify dissipation in a single trajectory or a single cell.

The joint probability distribution, $\phi(i,Q,t|i_0,0,0)$, (denoted as $\phi_i(Q,t)$ hereafter) describes the conditional probability of the system to be at the state $i$ at time $t$, after receiving $Q$ amount of medium entropy from the reservoir in the time interval $t$, starting at $t=0$, at the initial state $i_0$ and a state of zero entropy exchange. $\phi_i(Q,t)$ follows the equation(26),

$$\frac{\partial \phi_i(Q,t)}{\partial t} = \sum_{j(\neq i)} \left[ w_i^j \phi_j(Q - \Delta s_i^j, t) - w_j^i \phi_i(Q,t) \right]$$

(4)

Eq (4) can be used to monitor the kinetics of entropy exchanges in individual cells in a cell population. I investigated dissipation limited situations where the entropy exchange required for carrying out the stochastic transition become restricted. I consider two scenarios: (i) the total amount of entropy ($E$) available for exchange with the reservoir is fixed. This represents a situation where the total amount of energy available for dissipation is fixed. (ii) $E(t)$ increases at a fixed rate which is lower than that required to maintain the probability current in the steady state of Eq. (1). The above constraints are imposed in the kinetics in the following manner. The system is not allowed to make a transition $j \to i$ if that requires crossing of the limit $E$, i.e., $\phi(i,Q,t|j,Q-Q',t-\tau)=0$ when $Q > E(t)$. However, if other reactions (say, $j \to k$) at the state $j$ satisfying $Q \leq E(t)$ are available, then the particular reaction $j \to i$ is replaced by one of those reactions. Thus, the dissipation limit $E$ imposes a reflecting boundary condition(27) at $Q=E$ in Eq. (4). It is possible to solve Eq. (4) under this boundary condition exactly semi-analytically for simple cases when $E$ is a constant, however, for large number of receptors and ligands or a time dependent $E(t)$, such calculations become intractable. A continuous time MC method, akin to the standard Gillespie method (28), was developed here to simulate stochastic trajectories in these cases.

**B. An exactly solvable case:** Consider a single TCR interacting with a pMHC molecule in the minimal model. The signaling kinetics then involves transitions between three different states representing the unbound TCR and pMHC (state $T_1$), the TCR-pMHC complex (state $T_2$), and, the activated TCR-pMHC complex (state $T_3$). The biochemical reactions are described by,

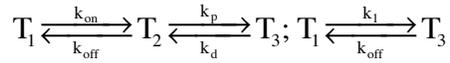

$$T_1 \underset{k_{off}}{\overset{k_{on}}{\rightleftharpoons}} T_2 \underset{k_d}{\overset{k_p}{\rightleftharpoons}} T_3; \quad T_1 \underset{k_{off}}{\overset{k_1}{\rightleftharpoons}} T_3$$

This simple example is amenable to analytical calculations and provides valuable insights into the kinetics in the dissipation limited case. The probabilities $p_1(t)$, $p_2(t)$, and $p_3(t)$, follow the equation,

$$\frac{dp_i}{dt} = \sum_{j=1}^{3} L_{ij} p_j \tag{5}$$

where,

$$L = \begin{bmatrix} -k_1 - k_{on} & k_{off} & k_{off} \\ k_{on} & -k_{off} - k_p & k_d \\ k_1 & k_p & -k_d - k_{off} \end{bmatrix}.$$

Exact solution of Eq. (5) (details in the web supplement) shows that at the steady state, $p_3(t\to\infty) \sim 1/(k_{\text{off}})^2$ for weak affinity ligands ($k_{\text{off}} \gg k_{\text{on}}$, $k_{\text{off}} \gg k_d$) and $k_p > k_d$. The mean value and higher moments of Q are usually calculated by solving the kinetics for the moment generating function, $\psi_i(\lambda,t) = \int dQ\, e^{\lambda Q} \phi_i(Q,t)$ (19, 26). $\psi_i(\lambda,t)$ follows the equation (19, 26),

$$\frac{\partial \psi_i(\lambda,t)}{\partial t} = \sum_j H_{ij}(\lambda) \psi_j(\lambda,t) \tag{6}$$

, where, $H_{ij}(\lambda) = (w_i^j)^{1-\lambda}(w_j^i)^{\lambda}(1-\delta_{ij}) - \delta_{ij}\sum_{j'\neq i} w_{j'}^i$. However, as shown in the web supplement, direct solution of Eq. (6) can be avoided and the moments of Q at all times can be recursively calculated analytically or semi-analytically using the solutions of Eq. (5). The calculations (details in the web supplement) show that the average rate of dissipation ($d\langle Q(t)\rangle/dt$) in the steady state is a constant and shows a peak at intermediate values of $k_{\text{off}}$ ($\sim \sqrt{k_d k_{on}}$). The ligand discrimination costs more energy at intermediate $k_{\text{off}}$ values because the system executes the KPR step more frequently compared to the low affinity or high affinity ligands. This also implies that the ligands with intermediate values of $k_{\text{off}}$ will arrive at a dissipation limit faster than ligands with other affinities. Eq. (4) with a reflective boundary condition at $Q=E=const$ was analyzed with two goals in mind: (i) Find the general structure of the equation (equivalent of Eq. (4)) that the system should satisfy under this condition. Such an equation can be further used to formulate a continuous time MC method (or Gillespie's method) to simulate stochastic trajectories in dissipation limited cases. (ii) Explore if any non-trivial behavior emerges even in this simple set up. Next, I outline the derivation of the equation followed by the joint probability distribution, $\phi_i(Q,t)$, with the reflective boundary condition at $Q(t)=E$ for the simple example described above. The full derivation is shown in the web supplement. The kinetics of $\phi_1(Q,t)$, $\phi_2(Q,t)$, and, $\phi_3(Q,t)$, according to Eq. (4), are given by,

$$\frac{\partial \phi_1(Q,t)}{\partial t} = k_{off}\phi_2(Q+\Delta_1,t) + k_{off}\phi_3(Q-\Delta_3,t) - (k_{on}+k_1)\phi_1(Q,t)$$

$$\frac{\partial \phi_2(Q,t)}{\partial t} = k_{on}\phi_1(Q-\Delta_1,t) + k_d\phi_3(Q+\Delta_2,t) - (k_{off}+k_p)\phi_2(Q,t)$$

$$\frac{\partial \phi_3(Q,t)}{\partial t} = k_1\phi_1(Q+\Delta_3,t) + k_p\phi_2(Q-\Delta_2,t) - (k_{off}+k_d)\phi_3(Q,t) \quad (7)$$

where, $\Delta_1 = \ln(k_{on}/k_{off})$, $\Delta_2 = \ln(k_p/k_d)$, and, $\Delta_3 = \ln(k_{off}/k_1)$. The reflecting boundary condition at $Q=E$, demands $\phi_i(Q>E,t)=0$ for all $i$'s. Since the reflection boundary condition does not lead to loss of any stochastic trajectory, the equation for $\phi_i(Q=E,t)$ at the boundary, $Q=E$, is obtained by using the conservation of total probability(27), $p_1(t)+p_2(t)+p_3(t)=1$, where, $p_i(t)=\Sigma_Q \phi_i(Q,t)$. The resulting equation showed that imposing the boundary condition for a system at a state ($j$, $Q$) at time $t$, is the same as setting the transition rates ($\{w^j_{i'}\}$) to zero when those transitions $\{j \to i'\}$ lead to the crossing of the dissipation limit at $Q=E$. It is straightforward to construct a continuous time MC method following Gillespie's algorithm(28) to simulate stochastic trajectories in this situation (see Materials and Methods section). An exact solution of Eq. (7) with reflective boundary conditions at two boundaries $Q=E_1$ and $Q=E_2$ was obtained by calculation of the eigenvalues and eigenvectors of the linear system (details in the web supplement). The comparison between the exact solution and the MC simulations showed an excellent agreement (Fig 1B).

**Results**

***A. Arrested states arise when dissipation is limited in the example of a single TCR interacting with a single pMHC.*** Analysis of Eq. (7) for the case of a single TCR and a single pMHC with a fixed dissipation limit at $Q=E$ shows the presence of arrested states in the kinetics, where, the system becomes confined to single state (e.g., state 1) or

multiple states (e.g., states 1 and 2) for a very long time ($\sim 1/k_1$). The specific nature of the arrested state and the time when it occurs depend on the values of the rate constants and the dissipation limit E. The physical origin of the arrested states is discussed below (Fig. 1C). Suppose, the system reaches the energy dissipation limit ($Q=E$) when it arrives at state 1 at time $t$. The value of $t$ will particularly depend on how often the KPR step was executed in the stochastic trajectory before it reached the limit, because, this step induces a non-zero probability current flow in the kinetics and its execution requires a much larger entropy flow ($\ln(k_{off}/k_1)$) compared to the other transitions for a biologically relevant model ($k_1 \ll (k_{off}, k_{on}, k_p, k_d)$). The possible transitions at state 1, 1→2 and 1→3, will need entropy flows, $\Delta s^1_2 = \ln(k_{on}/k_{off})$, and, $\Delta s^1_3 = \ln(k_1/k_{off})$, respectively, to the system from the reservoir. Since, $k_1 \ll k_{off}$, $\Delta s^1_3 < 0$, the system can release entropy to the reservoir and move below the dissipation limit $E$ by executing the reaction 1→3. However, this reaction (1→3) occurs within a time scale of $1/k_1$ which can be much longer than any time scale of biological or physical interest. Different kinetic responses arise for high affinity, and, moderate and low affinity ligands in this long time interval ($\sim 1/k_1$) (Fig. 1B). <u>*High affinity ($k_{off} \leq k_{on}$) ligands*</u>. Since, $\Delta s^1_2 > 0$, the transition 1→2 cannot occur without crossing the limit at Q=E. Thus, the system will remain at state 1 for a time scale of $1/k_1$. <u>*Moderate and low affinity ($k_{off} > k_{on}$) ligands.*</u> $\Delta s^1_2 < 0$, thus, the transition 1→2 could occur without crossing the limit at $Q=E$. However, after the system reaches state 2, the possible transitions, 2→3 and 2→1, are associated with with entropic flows, $\Delta s^2_3 = \ln(k_p/k_d) > 0$ (since $k_p > k_d$) and $\Delta s^2_1 = -\Delta s^1_2 = -\ln(k_{on}/k_{off})$, respectively. When, $\Delta s^2_3 \leq \Delta s^1_2$ or $\ln(k_p/k_d) \leq \ln(k_{on}/k_{off})$ or $k_{off} \leq k_{on}(k_d/k_p)$, the transition to 2→3 can occur without breaching the limit $Q=E$ and the system stays mobile between the states 1,2, and 3, without executing the the KPR step (3→1). However, when $\Delta s^2_3 > \Delta s^1_2$ or $k_{off} > k_{on}(k_d/k_p)$, the entropy gain from the previous 1→2 transition is not sufficient to support the 2→3 transition but can support the 2→1 transition, as a result, the system becomes confined between the states 1 and 2. The above described properties of the kinetics are also present when the system reaches the dissipation limit precisely or there is a small gap (e.g., $E$-$Q$<min($|\Delta_1|$, $|\Delta_2|$, $|\Delta_3|$)) between the dissipation limit and Q.

How do the properties of the arrested and the mobile states change when there are multiple ligand and receptor molecules? I investigated this question the next using MC simulations of Eq. (4) (see Materials and Methods for details). Interestingly, the results showed that the kinetics in the KPR model in these situations are similar to that of tagged molecules in models of glass formers at low temperatures marked by temporal clustering of stochastic events describing transitions between mobile and immobile states or dynamic facilitation(23).

***B. Kinetics with a fixed dissipation limit displays dynamic facilitation for multiple TCRs interacting with multiple pMHCs.*** First, I studied the kinetics in the presence of a fixed dissipation limit at $Q=E=const.$. MC simulation of stochastic trajectories showed three key differences with its counterpart without the dissipation limit (Fig. 2A). (i) The kinetics slowed down substantially once the system reaches the dissipation limit. (ii) Presence of large copy number fluctuations (Fig. 2A and Fig. S1). (iii) Stochastic events in the neighborhood of low and high activation states appeared to be bunched in time. Since, similar features are also observed in kinetics of tagged molecules in models of glass formers below the glass transition temperature, I analyzed the above features further by calculating quantities that are frequently used in characterizing kinetics in glass formers(23). A CTRW representation is often used for analyzing the glassy kinetics of the tagged molecules in glass formers(29), I will use a similar analogy for studying the stochastic trajectories in the minimal model.

*1. Analysis of the kinetics using a CTRW representation*: As described in Model section, the stochastic kinetics in the minimal model can be represented by a CTRW model where the walker moves on a two dimensional square lattice spanned by $N_{TM}$ and $N_{T*M}$ with a waiting time distribution, $P_w(\tau)$. The waiting time, $\tau$, is the time the random walker stays put at any state before making the next jump. The CTRW representing kinetics of tagged probes in glass formers is marked by a decoupling between diffusion and relaxation time scales(30, 31). The relaxation time scale is proportional to the persistence time scale ($t_p$) defined as the duration an initial state does not change(30, 31). The diffusion time scale is related to the exchange time ($t_x$), the time interval between any two subsequent random steps in the CTRW model(30, 31). The distributions of the time scales, $t_p$, and, $t_x$, show different forms ($P(t_p) \neq P(t_x)$) for the tagged molecules in glass formers due to the disparity between the time scales(23, 30, 31). The kinetics in glass formers also display ergodicity breaking where the time average of an observable over a long time interval is not equal to the ensemble average(23, 24). In a CTRW model, this can be caused by a power-law variation of $P_w(\tau)$(24) or when the waiting times in the subsequent steps become correlated(32). For example, the kinetics of potassium channels in the plasma membrane show a power law waiting time distribution(33) or dynamics of financial markets show correlation in successive waiting times(24, 32). In order to probe the presence of ergodicity breaking and its underlying origin in the kinetics of the minimal model, I calculated $P_w(\tau)$ and the correlation ($C(n)$) between subsequent waiting times. $C(n)$ is defined as,

$$C(n) = \langle 1/M \sum_{m=1}^{M} \tau_m \tau_{n+m} \rangle - \langle 1/M \sum_{m=1}^{M} \tau_m \rangle^2, \tag{8}$$

where, $\tau_n$ represents the waiting time at the nth step taken by the random walker. A finite $C(n)$ for $n \neq 0$ would indicate temporally correlated movements (32) or dynamic facilitation. The ergodicity breaking is characterized by calculating the ensemble averaged ($\langle r^2(t) \rangle$), defined as(24),

$$\langle r^2(t) \rangle = \langle (m-m_0)^2 \rangle + \langle (n-n_0)^2 \rangle, \tag{9}$$

where, the random walker starts at the position $\vec{r}(0) \equiv (m_0, n_0)$ at time $t=0$ and reaches at $\vec{r}(t) \equiv (m,n)$ at time $t$, and $\langle \cdots \rangle$ denotes average over an ensemble of trajectories. $\langle r^2(t) \rangle$ is compared with the time averaged mean squared distance for a single stochastic trajectory(24) defined as,

$$\overline{\delta^2(t,T)} = 1/(T-t) \int_0^{T-t} d\tau (\vec{r}(t+\tau) - \vec{r}(\tau))^2. \tag{10}$$

The presence of (weak) ergodicity breaking implies, when $T \gg t$(24),

$$\overline{\delta^2(t,T)} \neq \langle r^2(t) \rangle \tag{11}$$

The extent of ergodicity breaking in the kinetics can be quantified further by the ergodicity breaking parameter(24), $EB$,

$$EB = \lim_{T \to \infty} \left[ \langle \xi^2 \rangle - \langle \xi \rangle^2 \right] \tag{12}$$

where, $\xi = \overline{\delta^2(t,T)} / \langle \overline{\delta^2(t,T)} \rangle$. $EB=0$, for a kinetics with ergodicity (such as the standard Brownian motion(30)), and, $EB>0$ when ergodicty is broken in the kinetics(24).

2. *Calculation of specific quantities in the CTRW*: The calculation showed an exponentially decaying $P_w(\tau)$ in the absence of any dissipation limit (Fig. S2). This is expected as, in this case, the waiting time, $\tau$, at any state in the CTRW is distributed exponentially with a mean value ($\mu$) equal to inverse of the sum of the propensities of the outgoing transitions(28). Thus, $P_w(\tau)$ for a time interval of t is given by the superposition of exponential distributions with appropriate weights $g(\mu)$, i.e., $\sum_{\mu_{min}}^{\mu_{max}} g(\mu) e^{-\mu\tau}$. When $g(\mu)$ does not change with $\mu$ appreciably, the smallest $\mu$ ($=\mu_{min}$) makes the largest contribution the sum producing an exponential form for $P_w(\tau)$. However, in the presence of the dissipation limit the distribution displayed a much slower decay (non-Debye) than the exponential decay. This can occur when $g(\mu)$ varies with $\mu$ with a particular form pertaining to hierarchically constrained dynamics(34). The slower decay of $P_w(\tau)$ in the dissipation limited case is a manifestation of increased occurrences of longer waiting times characterizing the slow kinetics. However, the non-Debye exponential form of $P_w(\tau)$ alone does not establish dynamic facilitation or ergodicity breaking in the kinetics.

Calculations showed $P(t_p)=P(t_x)$ in the absence of the dissipation limit demonstrating the equivalence between the time scales $t_p$ and $t_x$ (Fig.2B, inset). Imposing the dissipation limit broke the equality (i.e., $P(t_p) \neq P(t_x)$) and both $P(t_p)$ and $P(t_x)$ displayed non-Debye decays, and, $P(t_p)$ agreed well with a stretched exponential decay ($\propto \exp(-at_p^\beta)$) for over 3 decades (Fig. 2B). $t_p$ is associated with the relaxation time scale of the initial state, and, in glass formers it corresponds to relaxation of spatial structures(23). Whereas, $t_x$ is associated with the diffusive time scale of the random walker. The emergence of the stretched exponential or non-Debye relaxation times in glassy systems are accompanied with hierarchical activation of underlying microscopic processes(23, 34). When the system reaches the dissipation limit, certain reactions can take place only when appropriate amount of entropy is released by a concerted execution of a series of reactions, this provides a source for hierarchical activation in the system. In the simulations, $\langle t_p \rangle$ is about three times larger than $\langle t_x \rangle$. Similar behavior ($\langle t_p \rangle > \langle t_x \rangle$) in glass formers indicates breakdown of the Stokes-Einstein relationship relating dissipative and diffusion timescales in liquids(23, 35). The non-equivalence of $t_p$ and $t_x$, as in glass formers, points to the presence of dynamic facilitation or clustering of mesoscopic events in time(23, 35).

The correlation function, C(n), as defined in Eq. (8), further characterized the nature of the dynamic facilitation in the KPR model. Calculation of C(n) showed that waiting times separated by multiple events are more correlated in the dissipation limited case compared to that with no dissipation limit (Fig. 2C). C(n) decreases substantially within a single step when there was no dissipation limit (Fig. 2C). Next, I investigated if these correlations are able to generate ergodictiy breaking as found in the models of CTRW with correlated time steps. The calculations of $<r^2(t)>$ (Eq. (9)) and $\overline{\delta^2(t,T)}$ (Eq. (10)) showed that $<r^2(t)> \neq \overline{\delta^2(t,T)}$ in the KPR model with the dissipation limit demonstrating a breakdown of ergodicity in the kinetics due to the confinement of stochastic trajectories in specific regions in the state space for very long times ($\sim 1/k_1$) (Fig. 2D and Fig. S9). $\overline{\delta^2(t,T)}$ saturates at large t as the values of $N_{TM}$ and $N_{T*M}$ are bounded by total numbers of T and M (see also Fig. S10). Removing the dissipation limited restored ergodicity (Fig. 2D and Fig. S9), i.e., $<r^2(t)> = \overline{\delta^2(t,T)}$. The ergodicity breaking is further quantified by calculating the ergodicity parameter, EB (Eq. (12)), for the KPR model with and without the dissipation limit. In the presence of the limit, non-zero EB values were generated (Fig. S9), whereas, in the absence of any limit, EB becomes vanishingly small (Fig. S9).

***C. Kinetics with a fixed rate of energy supply.*** This case was investigated by including a variable in the biochemical reactions that increased the medium entropy (or the available energy for dissipation) in the reservoir at a constant rate ($e_r$) (see Materials and Methods for details). The simulations were performed for the cases where the required rate of energy dissipation (v) was larger than that available from the reservoir. Analysis of the kinetics revealed the presence of two dynamically distinct regions (Fig. 3A). (i) For times $0 < t \leq \tau_{trans}$, most of the medium entropy produced in the reservoir flowed into the system to fuel the reactions. At the end of $\tau_{trans}$, when the total medium entropy inflow into the system became comparable to the medium entropy required to bring the initial state to the steady state of the NKPR model (Fig. S3), the kinetics moved into the second regime. (ii) For $t > \tau_{trans}$, the entropy received by a single cell (or single trajectory) did not change appreciably over a time scale $\tau_{diss}$ despite medium entropy being produced in the reservoir. Beyond $\tau_{diss}$, the KPR step is executed and the medium entropy flow into the system changes abruptly by $\sim \ln(k_{off}/k_1)$. A possible mechanism explaining the above behavior is when $t \leq \tau_{trans}$, the medium entropy produced in the reservoir is fully spent on carrying out the biochemical reactions, however, since the KPR step requires the largest amount of entropy influx ($\sim \ln(k_{off}/k_1)$) it rarely takes place in this regime, and consequently, the system evolves as the dissipation limited NKPR model. Towards the end of $\tau_{trans}$, when the stochastic trajectories in the system are close to the dissipationless steady state of the NKPR model, the system does not draw much medium entropy from the reservoir over a time scale of $\tau_{diss}$. This results in accumulation of sufficient medium entropy in the reservoir to fuel the execution of the KPR step at the end of $\tau_{diss}$ (or $\tau_{diss} e_r \geq \ln(k_{off}/k_1)$). Thus, the time evolution for $t > \tau_{trans}$ can be intuitively thought of as successions of time segments of scale $\tau_{diss}$ where the kinetics is similar to that of the dissipationless steady state of the NKPR model. Further analysis of the simulation results confirmed the above picture.

Calculation of $P(t_p)$ and $P(t_x)$ showed that for $t < \tau_{trans}$, $P(t_p) \neq P(t_x)$, suggesting similarities of the kinetics to that of the fixed dissipation limit case (Fig. 3B). For $t > \tau_{trans}$, I found $P(t_p) \approx P(t_x)$, and both the distributions decayed exponentially as in unlimited dissipation cases (Fig. 3B, inset). Distributions of $N_{T^*M}$ and $N_{TM}$ demonstrated that the system closely follows the steady state of the NKPR model for $t > \tau_{trans}$ (Fig. S4). The value of $\tau_{trans}$ is roughly related to $e_r$, the available medium entropy ($Q_0$) at t=0, and, the total medium entropy required to change the initial state to the steady state of the NKPR model ($Q^{NKPR}_{steady}$) as, $\tau_{trans} \approx (Q^{NKPR}_{steady} - Q_0)/e_r$. C(n) and $\overline{\delta^2(t,T)}$ showed the emergence (or absence) of dynamic facilitation and ergodicity breaking for $t < \tau_{trans}$ ( or $t > \tau_{trans}$) (Figs. S5 and S6). The difference between $\langle t_p \rangle$ and $\langle t_x \rangle$ calculated at increasing values of $e_r$ showed that, for $t < \tau_{trans}$, increasing $e_r$ decreased the magnitude of the difference which reaches zero as $e_r$ increases to $e_r \geq v$ (Fig. S7). Similarly, dynamic facilitation and ergodicity breaking disappears at $t < \tau_{trans}$ for $e_r \geq v$. Thus, at $t < \tau_{trans}$ increasing $e_r$ appears to generate a qualitatively similar effect as increasing the temperature across the glass transition in glass formers.

***D. Implications for ligand discrimination.*** The emergence of glassy kinetics in the dissipation limited negatively affects ligand discrimination. Arrested states slow down the kinetics in addition to making an undesired state (e.g., TM for low $k_{off}$) persist in single cells over a long time scale ($\sim 1/k_1$ for fixed dissipation limit or $\tau_{diss}$ for a fixed rate of entropy increase). Both these effects oppose a successful discrimination program. Without the dissipation limit, the biochemical kinetics reached the steady state in a short time scale (~mins), where the cell population average of the activated species (T*M) decreased with the ligand affinity $\langle N_{T*M} \rangle \sim 1/k_{off}^2$ allowing the cells to discriminate between pathogenic (low $k_{off}$) and self ligands (high $k_{off}$) with a greater sensitivity (Fig. 4). Limiting dissipation qualitatively changed this pattern where $\langle N_{T*M} \rangle$ displayed a non-monotonic variation with $k_{off}$ at short times (~mins) (Fig. 4). In this case, decreasing ligand affinity leads to an increase in the activation producing an outcome opposite to that required by a successful discrimination program. When energy for dissipation is supplied at a fixed rate, for time scales $t < \tau_{trans}$, the response is similar to that with a fixed dissipation limit, and, at longer time scales ($t > \tau_{trans}$), the system responds with a lower precision ($\langle N_{T*M} \rangle \sim 1/k_{off}$) compared to the unrestricted KPR model. Thus, the response at long time scales in this case is similar to that of the less discriminatory NKPR model. However, depending on the initial (basal) signaling state of the single cells, $\tau_{trans}$ could be much longer than biologically relevant time scales (~mins).

Moreover, in a dissipation limited scenario, large cell-to-cell variations of copy numbers of activated species will hinder discrimination when multiple types of ligands are presented simultaneously to a T cell population. For example, a successful discrimination program requires that T cells should be able to recognize a small fraction ($f_{path}$) of pathogenic ligands (say, $k_{off}= k_{path}$) within a large population ($f_{self} \gg f_{path}$, and, $f_{self}+ f_{path}=1$) of self-ligands ($k_{off}= k_{self}$). Therefore, the discrimination program should generate widely different distributions (or P($k_{off}$, $N_{T*M}$)) of the active species (T*M) in a T cell population when ligands are presented with an input distribution,

$$P_{ligand}(k_{off}) = f_{path}\delta_{k_{off},k_{path}} + f_{self}\delta_{k_{off},k_{self}} \text{ or } P_{ligand}(k_{off}) = \delta_{k_{off},k_{self}}$$

. $f_{path}(f_{self})$ denotes the fraction of pathogenic (self) ligands presented to the T cells. The large variation in $N_{T*M}$ in the dissipation limited cases will produce a wide spread in P($k_{off}$, $N_{T*M}$) (Fig. S8). Consequently, there will be a substantial overlap between the above input distributions leading to a much poorer discrimination (Fig. S8) in the dissipation limited case.

**Discussion**

The analysis carried out here showed that restricting energy dissipation qualitatively changes signaling kinetics of high precision responses functioning outside thermodynamic equilibrium. The changes are marked by advent of slow kinetics, long-lived arrested states, dynamic facilitation, and, ergodicity breaking. The origin of this emergent behavior is purely dynamical and arises due to dynamical constraints imposed by limited dissipation. The appearance of the glassy kinetics rectifies the naïve intuition that in the presence of a dissipation limit the system will fall back its dissipationless counterpart (e.g., the steady state kinetics without the KPR step). The results show, in contrast to the naïve intuition, when the energy for dissipation is limited by a fixed amount, the kinetics becomes confined to specific biochemical states for long durations, and, when energy for dissipation is supplied at a fixed rate, depending on the energy supply rate and the initial state of the system, the kinetics for a long time can behave similar to that with a fixed dissipation limit. Furthermore, the breakdown of ergodicity in the dissipation limited cases points to a basic difficulty in deriving details regarding single cell kinetics from cell population data.

The emergence of the glassy kinetics, characterized by long-lived states (activated or de-activated) and large cell-to-cell variations of copy numbers of signaling products, prove to be detrimental to the discrimination program involving KPR. KPR is a central concept used by biophysical models(2, 15, 21) describing experiments pertaining to ligand discrimination in immune cells, such as T cells. The presence of the KPR step in these models leads to complete or partial reversal of intermediate activated states breaking the detailed balance condition(8, 18, 20). As a result, these models work outside thermodynamic equilibrium and a constant probability current in the network in the steady state is sustained by a constant supply of energy. Therefore, the qualitative features (e.g., slow kinetics, large cell-to-cell variations, ergodicity breaking, poor ligand discrimination) that arise due to limiting dissipation in the KPR scheme are likely to impact the detailed biophysical models of ligand discrimination the available energy becomes restricted. A possible test of these results will involve single cell experiments carried out in energy limited conditions, possibly induced by manipulation of nutrient metabolism or signaling events regulating ATP production(9).

It is assumed in the minimal model that molecules are well mixed in the simulation volume describing a small region (1μm$^2$ area x 0.01μm depth) proximal to the plasma membrane. However, immune receptors and associated signaling molecules can be distributed inhomogeneously in larger regions of the plasma membrane (e.g., microclusters)(36, 37) and in the cytosol(38). The kinetics of single molecules in the spatially heterogeneous cellular environment displays ergodicity breaking in certain biological systems(24, 33, 39). For example, spatial kinetics of single potassium channels in the plasma membrane was observed to follow a CTRW model with a power law waiting time distribution that led to weak ergodicity breaking in the kinetics(33). In contrast, in the minimal model, the biochemical reaction kinetics showed ergodicity breaking that arose due to finite correlations between successive waiting times. It will be an interesting future direction to study the interplay between the diffusion and the reaction kinetics where ergodicity breaking in the two types of kinetics could be induced by spatial heterogeneity and dissipation limitation, respectively.

The framework considered here for quantifying dissipation does not explicitly include activation energy(40). Thus, if the system resides at the free-energy limit (i.e., Q=E) and a particular reaction (say, 1→2) releases medium entropy to the reservoir (e.g., $\Delta s_m{}^1{}_2$ =ln($w^1{}_2$/ $w^2{}_1$) <0), the reaction is then assumed to occur. However, it is possible that the reaction also requires crossing of an activation barrier(40) and thus might not occur in this situation. This would impose a stricter restriction on the reactions that can potentially arise at the dissipation limit. Therefore, realistically there could a larger number of arrested states and a greater degree of dynamic facilitation in the signaling kinetics.

The kinetics in the dissipation limited cases in the KPR model demonstrates similarities with that in glass formers in terms of the appearance of slow kinetics and dynamic facilitation. However, the glassy kinetics in the two systems also shows few important contrasts. For example, the shapes of $P(t_p)$ and $P(t_x)$(23) are different in these models. In glass formers, the glassy kinetics arises when the temperature is lowered past the glass transition temperature and the system undergoes a phase transition in the space and time of stochastic trajectories(23, 41). In the KPR model, the notion of temperature or any phase transition is not evident. In simple networks violating detailed balance reveal dynamical phase transitions between localized and de-localized states induced by increasing entropy production rate in the limit of large system sizes (42). Increasing the rate ($e_r$) of medium entropy supply in the minimal model produces changes in the kinetics like the temperature, however, further work is required to make this connection transparent or establish any presence of a phase transition in the KPR model.

**Materials and Methods**
*MC simulations:* A continuous time MC method was used to simulate stochastic trajectories in the minimal model for multiple receptor and ligands. The construction of the Master equation for $\phi_i(Q,t)$ for the dissipation limited case shows that the propensities ($\{w_j^i\}$) of the reactions that take the system over the dissipation limit ($Q(t)=E$)) should be set to zero.  This result was used to construct a Gilliespie(28) like algorithm by calling two uniform random numbers ($r_1$ and $r_2$) in the unit interval for simulating the trajectories. In the simulations, a variable, $Q(t)$, keeps track of the entropy that flows into the system from the reservoir. The time for the next reaction ($\tau$) in the system residing at state i at time t with a total entropy exchange $Q(t)$ is given by $\tau=1/a_{total} \ln(1/r_1)$, where the total propensity, $a_{total}=\Sigma_j w_j^i$. The next reaction ($i \to \mu$) is chosen by calculating $\mu$ satisfying the condition, $\sum_{j=1}^{\mu-1} w_j^i \leq a_{total} r_2 < \sum_{j=1}^{\mu} w_j^i$. Any propensity ($w_j^i$) that results in $Q(t)+\ln(w_j^i/w_i^{j'})>E$ is set to zero for the calculations of $\tau$ and $\mu$ in the above steps. $Q(t)$ is updated to $Q(t+\tau)=Q(t)+\ln(w_\mu^i/w_i^\mu)$ after the transition $i \to \mu$ is executed. When there is a supply of medium entropy at a constant rate ($e_r$) in the reservoir, a stochastic variable $q$, decoupled from rest of the variables in the minimal model, is introduced. q increases by unity ($q \to q+1$) with a propensity $w_{q+1}^q = e_r$, increasing the reservoir medium entropy (i.e., $E(t+\tau) \to E(t)+1$) by unity at every execution. $w_{q+1}^q$ is used along with other propensities in the model for the calculations of $\tau$ and $\mu$. The condition, $Q(t)+\ln(w_j^i/w_i^{j'})>E(t)$ is used to set a reaction ($i \to j'$) propensity that crosses the dissipation limit to zero. $w_{q+1}^q$ is not considered in evaluation of the above condition and is never set to zero.

*Calculations of* (i) $P_w(\tau)$, (iia) $P(t_p)$, (iib)$P(t_x)$, (iii)$C(n)$, *and*, (iv) $\overline{\delta^2(t,T)}$ : (i)Waiting times ($\tau$) in a time interval $T(\gg \tau)$ were calculated for each stochastic trajectory in the CTRW model and $P_w(\tau)$ was calculated using all the $\tau$'s collected in a large ensemble of stochastic trajectories. (iia,b) A start time $t_{start}$ was chosen. If the next reactions in a CTRW stochastic trajectory occurred at times, $t_1, t_2, t_3, \cdots$, then $t_p$ for the trajectory was defined as, $t_p=t_1-t_{start}$, and, $t_x$ was calculated using $t_2-t_1$, $t_3-t_2$, and, so on(31). $t_{start}$ was chosen at times after the system reached the dissipation limit for the fixed dissipation limit. When the dissipation limit increases with a rate $e_r$, $t_{start}$ was chosen either at $t<\tau_{trans}$ or $t>\tau_{trans}$. Values of $t_p$ and $t_x$ were collected over a large number of stochastic trajectories ($>10^4$) for the evaluating $P(t_p)$ and $P(t_x)$. (iii) The reactions that take place after time $t_{start}$ are indexed as 1, 2, ..., n+1 at times, $t_1, t_2, t_3, \cdots, t_{n+1}$, respectively. The waiting times $\tau_1$ (=$t_2-t_1$), $\tau_2$ (=$t_3-t_2$), .., $\tau_n$ (=$t_{n+1}-t_n$) were used to calculate C(n) using Eq. (8) for a large ensemble of stochastic trajectories. For the fixed dissipation limit, I set $t_{start}=0$, and, when the dissipation limit increases with a rate $e_r$, $t_{start}$ was chosen either at $t<\tau_{trans}$ or $t>\tau_{trans}$. (iv) A stochastic trajectory in the CTRW model, simulated for a long time $T(\sim 2\times 10^4 s)$, was assigned positions ($r(t_i)=(m_i,n_i)$) at regular time intervals of $\Delta t$ ($t=\{t_1,\ldots,t_N=T_N\}$). The time averaged $\overline{\delta^2(t,T)}$ was calculated by replacing the integral in Eq. (10) by

$$\overline{\delta^2(n\Delta t, T_N)} = 1/(N-n)\sum_{i=1}^{N-n} (\vec{r}(t_i+n\Delta t)-\vec{r}(t_i))^2 \quad \overline{\delta^2(t,T)},$$ calculated for different stochastic trajectories, are shown in Fig. 2 and 3.

*Model parameters*: The simulation box in the MC simulations represents a region containing the plasma membrane (area=1μm$^2$) and a thin layer (depth=0.01μm) of the cytosol underneath the membrane. The size of the region was chosen such that the reaction time scales are larger than that of the diffusion time scales(43) (diffusion constant ~ 0.1-0.01 μm$^2$/s), thus, the molecules in the box can be considered well mixed. The data shown in the main text were carried out for the following values of the parameters: $k_{on}=0.003s^{-1}$, $k_p=1.0s^{-1}$, $k_d=0.1s^{-1}$, $k_1=10^{-8}s^{-1}$, $k_{off}$ is varied between $0.001s^{-1}$ to $10s^{-1}$, $N_{T0}=N_{M0}=100$ molecules/μm$^2$. The values are based on their measured values published in the literature. Details are provided in Table S1 in the web supplement. All the simulations were started off with $N_{TM}=N_{T*M}=0$.

**Figure Captions**

**Figure 1.** (A) Schematic diagram displaying the biochemical reactions in the minimal model. The KPR step is shown in green. The transition T+M→T*M (dotted line), occurring with a much smaller rate than the rest of the reactions, is assumed to keep the entropy calculations finite. (B) The exact solution (solid line) with two boundary conditions at E=3 and E=1 for $N_{T0}=N_{M0}=1$, $k_{on}=1/e$, $k_{off}=1/e^2$, $k_p=1$, $k_d=1/e$, $k_1=1/e^2$, $k_{off}'=1/e$ (rate for the KPR step), is compared with the developed continuous time MC scheme for $p_2(t)$ (∘) and $p_3(t)$(□). (C) Schematic diagrams showing the arrested and mobile states in the case with a TCR(T) interacting with a single pMHC(M) molecule. The arrows indicate the states where the system arrives at the dissipation limit.

**Figure 2.** (A) Stochastic trajectories obtained from the MC simulation for the copy numbers of T*M ($N_{T*M}$) in the presence and absence of the fixed dissipation limit (E=500). The parameters are set to values shown in the Materials and Methods section with the $k_{off}$=0.001s$^{-1}$. (B) Variation of P($t_p$) and P($t_x$) with their arguments for the dissipation limited case. The solid line shows a fit to P($t_p$) with a stretched exponential function (f(t)=5.17exp(-at$^\beta$), a=3.418, β=0.5622). (Inset)The differences between P($t_p$) and P($t_x$) disappear when there is no restriction for energy dissipation. Both the distributions decay exponentially. The parameters are the same as in (A). (C) Variation of C(n) with n for the dissipation limited and the unrestricted case. C(n) is scaled with C(0) to bring both the data on the same scale. The inset shows a close up of the main figure at smaller values of n. C(n) reaches 1/3 of C(0) in n≈10 when the dissipation is limited, whereas, for the case with unlimited dissipation, C(n) falls much below C(0)/3 at n=1. The parameters are the same as in (A). (D) Variation of $\overline{\delta^2(t,T)}$ with t for 20 different stochastic trajectories for the dissipation limited and the unlimited case (inset). The parameters are the same as in (A) with T=10$^6$s. The spread in $\overline{\delta^2(t,T)}$ for different configurations indicate ergodicity breaking which disappears when the dissipation is unlimited (inset). Over 10$^4$ trajectories were used for the all calculations above.

**Figure 3.** (A) Kinetics of the total amount of medium entropy influx $Q(t)$ into the system and the total amount of medium entropy produced in the reservoir $E(t)$ corresponding to a single stochastic trajectory or a single cell. The parameters are the same as that given in the Materials and Methods section, and, $k_{off}$=0.001s$^{-1}$ and $e_r$=0.01s$^{-1}$. For $t<\tau_{trans}$ all the produced medium entropy is consumed by the system, and, for $t>\tau_{trans}$, medium entropy accumulates in the reservoir for a time scale of $\tau_{diss}$. The corresponding kinetics for $N_{T*M}$ is shown in the inset. (Inset) The kinetics of $N_{T*M}$ for the NKPR model is shown for comparison. (B) Variation of P($t_p$) and P($t_x$) with their arguments for data collected at a time t (=20,000s <$\tau_{trans}$). The solid line shows a fit to P($t_p$) with a stretched exponential function (f(t)=14.4913exp(-at$^\beta$), a=10.543, β= 0.836). (Inset) The differences between P($t_p$) and P($t_x$) disappear at a later time t=50,000s ≫ $\tau_{trans}$). The solid line shows a fit

close to an exponential decay($\propto \exp(-x^{1.045}/0.0535)$). The parameters are the same as in (A). Over $10^4$ trajectories were used for all the calculations above.

**Figure 4**. Variation of $N_{T*M}$, averaged over a population of single cells (n=10,000) at t=5 mins, with $k_{off}$. The data are shown for the cases of a fixed dissipation limit (E=500), a fixed rate of medium entropy production ($e_r=0.01s^{-1}$), unlimited dissipation, and, the NKPR model. The dissipation limited cases offer a poorer discrimination with a decreased range of variation of $\langle N_{T*M}\rangle$ and a non-monotonic variation with $k_{off}$. To illustrate, T cells following the KPR model are able to discriminate between the ligand affinities (shown in dashed and dotted vertical lines) by crossing the activation threshold (horizontal solid line) for the stronger ligand, however, limiting dissipation abrogates this discrimination.

**Acknowledgement:** JD thanks Veronica Vieland, Susan Hodge, and Sang-Cheol Seok for helpful discussions. JD also thanks Ashok Prasad for a stimulating discussion. This work was partially supported by the W. M. Keck Foundation and the Research Institute at the Nationwide Children's Hospital.

**Competing Interests:** I have no competing interests to declare.

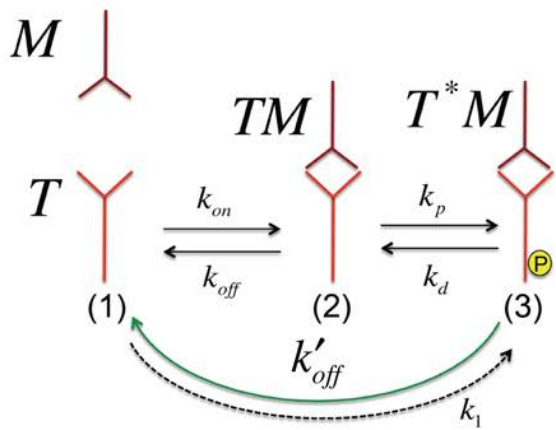
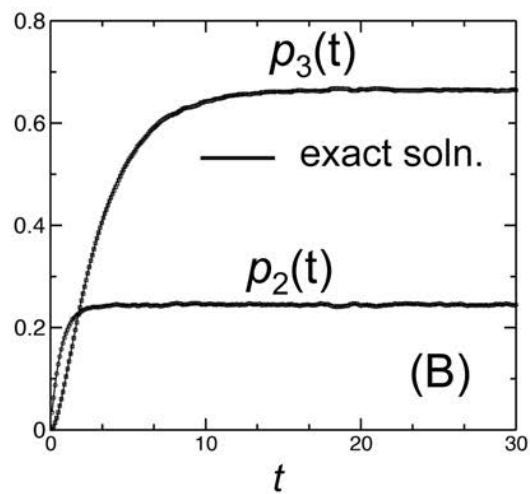
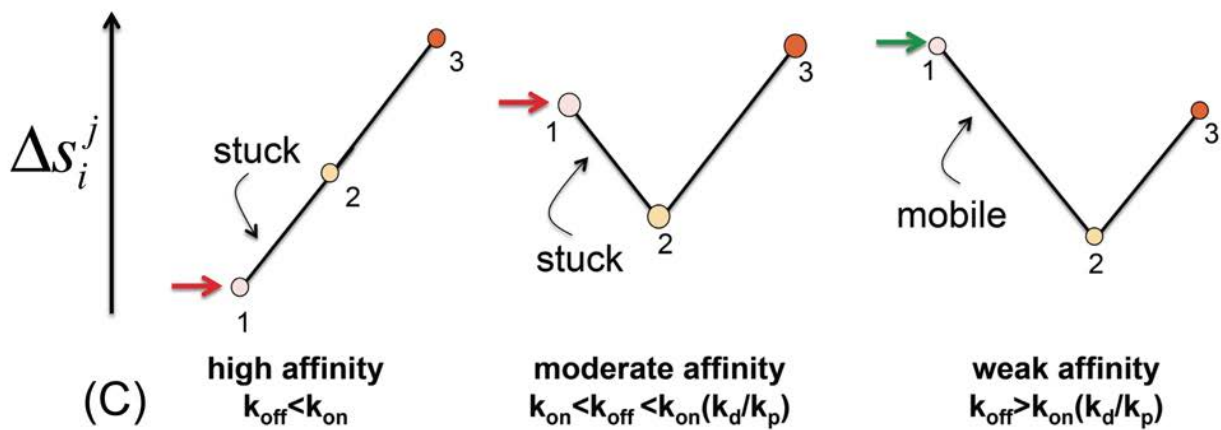

(A) Reaction scheme. (B) $p_2(t)$ and $p_3(t)$ exact solution. (C) Three affinity regimes: high affinity $k_{off} < k_{on}$ (stuck); moderate affinity $k_{on} < k_{off} < k_{on}(k_d/k_p)$ (stuck); weak affinity $k_{off} > k_{on}(k_d/k_p)$ (mobile).

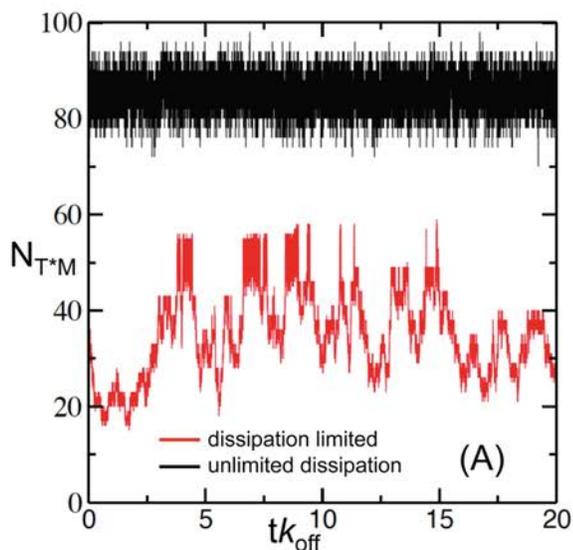
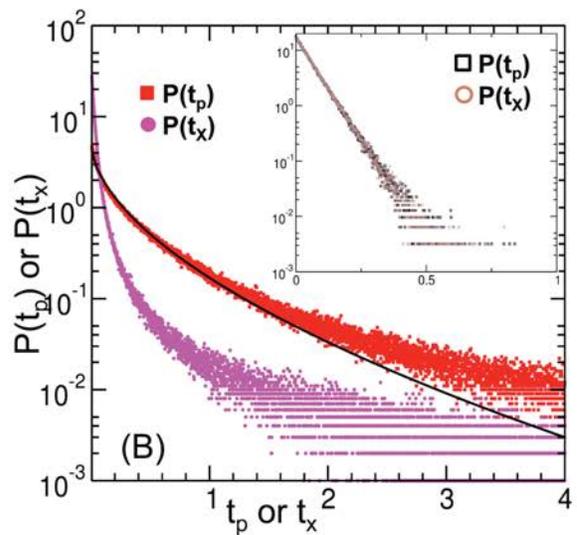
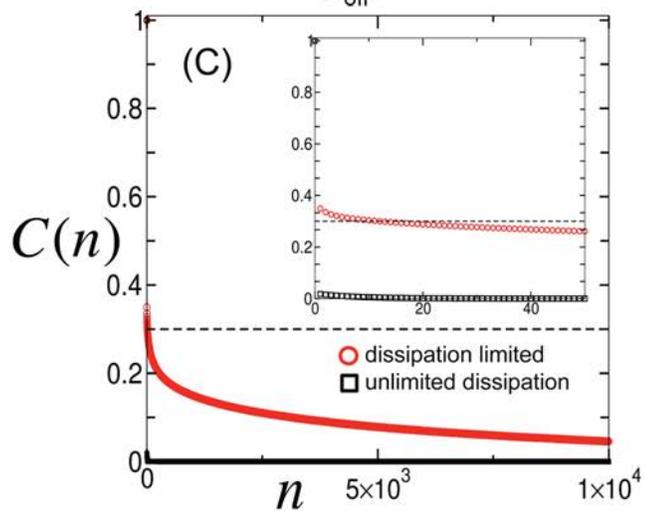
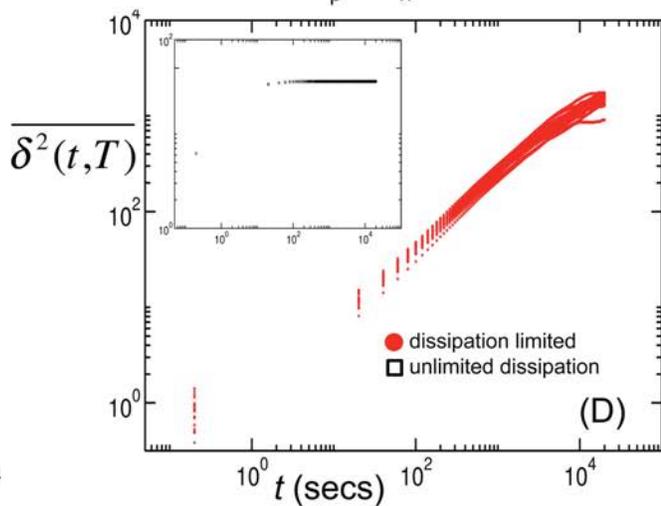

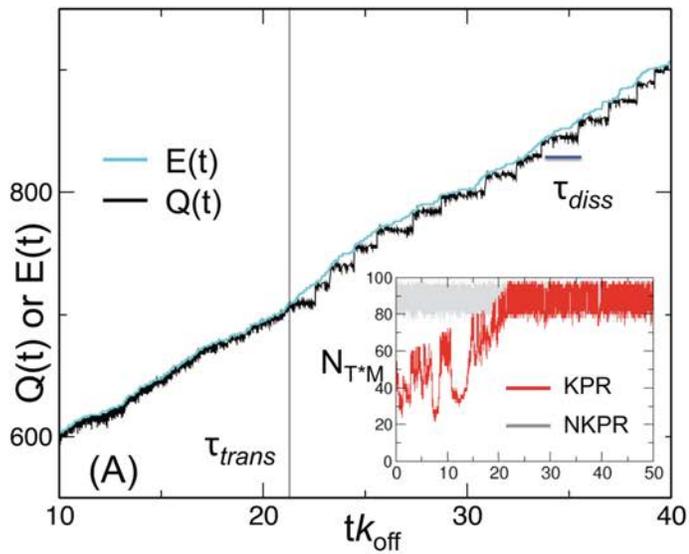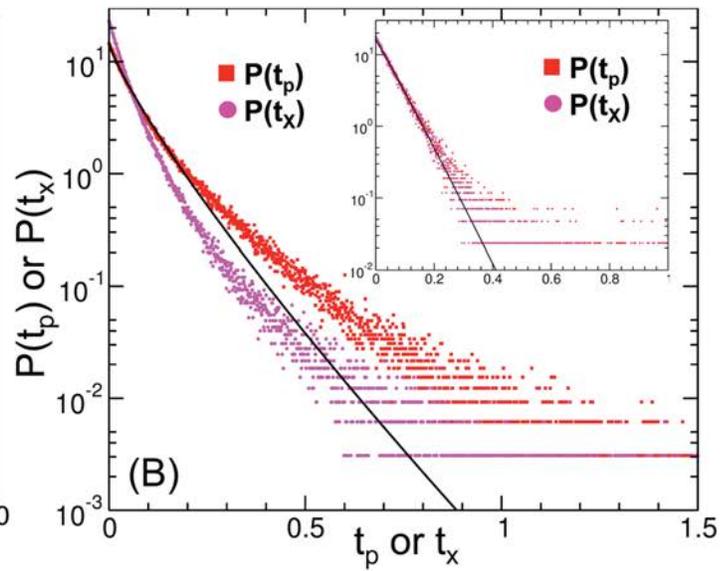

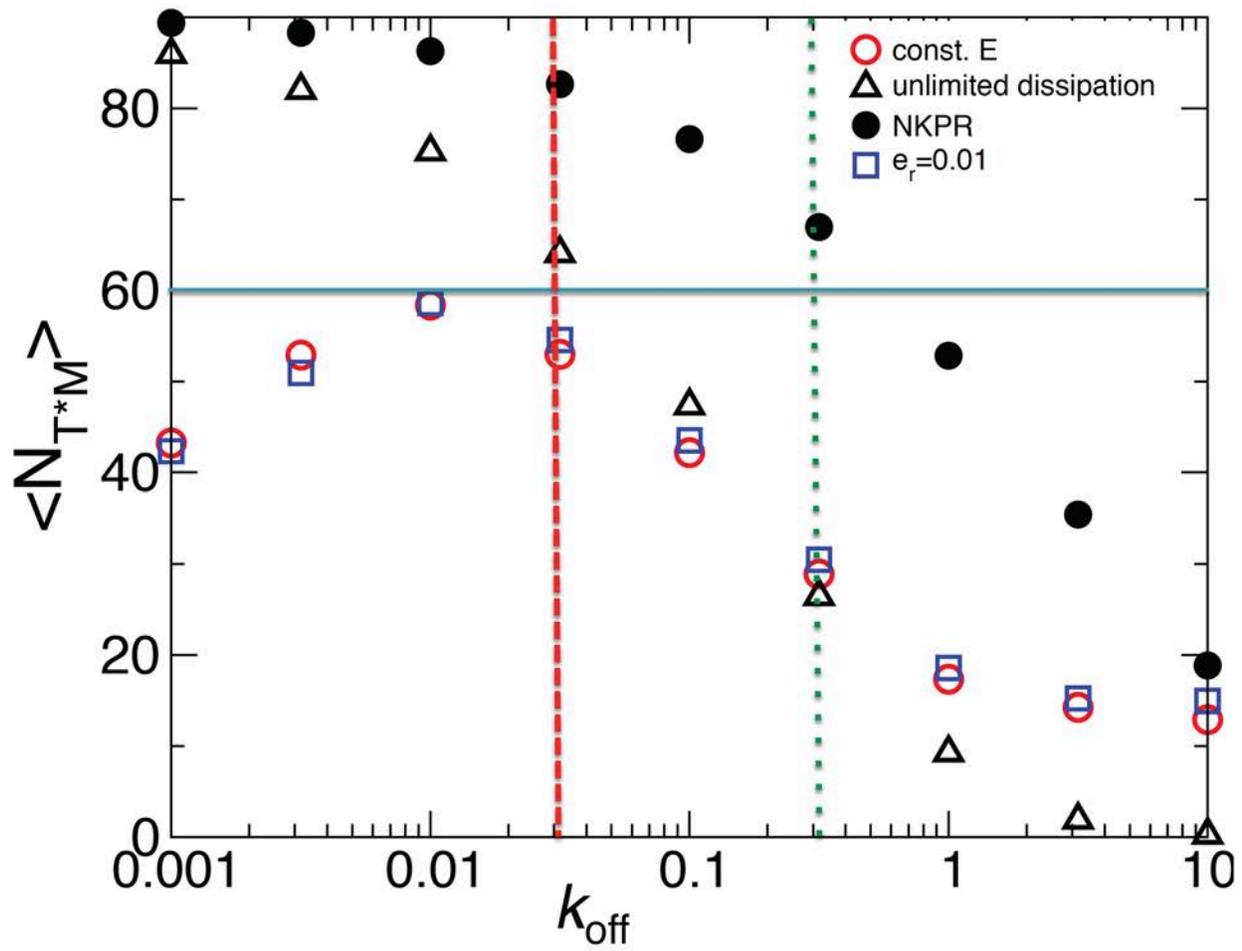

**Supplementary Material for "Limiting energy dissipation induces glassy kinetics in single cell high precision responses"**

## I. Exact Solution for the one receptor one ligand case

The kinetics is described by the reaction scheme,

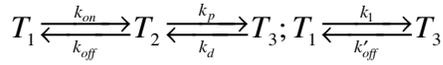

$$T_1 \underset{k_{off}}{\overset{k_{on}}{\rightleftharpoons}} T_2 \underset{k_d}{\overset{k_p}{\rightleftharpoons}} T_3; \; T_1 \underset{k'_{off}}{\overset{k_1}{\rightleftharpoons}} T_3$$

If the probabilities for being at the states 1,2, and, 3 are given by, $p_1$, $p_2$, and, $p_3$, then the corresponding Master equation is given by,

$$\frac{dp_i}{dt} = \sum_{j=1}^{3} L_{ij} p_j \quad (S1)$$

where,

$$L = \begin{bmatrix} -k_1 - k_{on} & k_{off} & k_{off} \\ k_{on} & -k_{off} - k_p & k_d \\ k_1 & k_p & -k_d - k_{off} \end{bmatrix}$$

I assumed $k_{off}' = k_{off}$ in Eq. (S1). One can scale the time with $k_p$ to render it dimensionless, e.g., $t = \tilde{t}/k_p$, then all the other rates can also be transformed into dimensionless variables, e.g., $\tilde{k}_1 = k_1/k_p$, $\tilde{k}_{off} = k_{off}/k_p$, and so on. For simplicity we do not display the tildes in the equation, thus, hereafter, all the rates will denote their dimensionless counterparts. The dimensionless, L, now takes the form,

$$L = \begin{bmatrix} -k_1 - k_{on} & k_{off} & k_{off} \\ k_{on} & -k_{off} - 1 & k_d \\ k_1 & 1 & -k_d - k_{off} \end{bmatrix}$$

The above equation can be easily solved. The solution for the initial condition, $p_1(0)=1$, $p_2(0)=p_3(0)=0$ is given by,

$$p_1(t) = \frac{e^{-(k_1+k_{off}+k_{on})t}(k_1+k_{on})+k_{off}}{k_1+k_{off}+k_{on}} \tag{S2}$$

$$p_3(t) = A/B$$

$$A = k_{on} + e^{-(k_1+k_{off}+k_{on})t}(k_1-1)(1+k_d+k_{off})(k_1+k_{on}) + e^{-(1+k_d+k_{off})t}(k_{on}-k_1k_d)(k_1+k_{off}+k_{on})$$

$$+(k_d+1-k_1)k_1(1+k_{off})+k_dk_{on}-k_1k_{on}(2+k_{off})-k_{on}^2$$

$$B = (1+k_d+k_{off})(k_1+k_{on}+k_{off})(1+k_d-k_1-k_{on})$$

$$p_2 = 1 - p_1 - p_3$$

At the steady state ($t \to \infty$),

$$p_1^s = \frac{k_{off}}{k_1+k_{off}+k_{on}}$$

$$p_3^s = \frac{k_{on}+k_1(1+k_{off})}{(k_1+k_{off}+k_{on})(k_d+k_{off}+1)} \tag{S3}$$

$$p_2^s = 1 - p_1^s - p_2^s$$

In the limit, $k_d \ll k_{off}$, $k_1 \ll (k_{off}, k_{on})$, $k_{off} < k_{on}$, and, scaling the parameters back with $k_p$,

$$p_3^s \to k_{on}k_p/k_{off}(k_p+k_{off}) \xrightarrow{k_{off}/k_p \gg 1} k_{on}k_p/k_{off}^2$$

When the KPR step is absent, $p_3^s \to k_{on}k_p/(k_dk_{off}) \propto 1/k_{off}$.

Thus, the presence of the KPR step leads to an increased sensitivity ($1/(k_{off})^2$ instead of $1/k_{off}$) of the steady state values of $p_3$ to changes in $k_{off}$ in the above range of parameters. Analysis of the steady states of the deterministic mass-action kinetics of the above system produces the same behavior.

**Estimation of dissipation**

The system entropy ($S_{sys}$) for the system is given by,

$$S_{sys}(t) = -\sum_{i=1}^{3} p_i(t) \ln p_i(t) \ .$$

The rate of change in $S_{sys}$ is given by,

$$\frac{dS_{sys}(t)}{dt} = -\sum_{\substack{i,j \\ j \neq i}}^{3} w_i^j p_j \ln\left(\frac{w_j^i p_i}{w_i^j p_j}\right) - \sum_{\substack{i,j \\ j \neq i}}^{3} w_i^j p_j \ln\left(\frac{w_j^i}{w_i^j}\right)$$

The first term ($dS_{total}/dt$) on the RHS is always non-negative and is associated with energy dissipation in the system. $w_i^j$ describes the rate of transition for the change, $j \to i$. The second term ($dS_{med}/dt$) gives the rate of entropy exchanged with the reservoir. Using the solution for Eq. (S1) the above rates can be easily calculated. At the steady state,

$$dS_{med}/dt|_{steady} = -\sum_{\substack{i,j \\ j \neq i}}^{3} w_i^j p_j \ln\left(\frac{w_j^i}{w_i^j}\right)$$

$$= \frac{1}{(k_1 + k_{off} + k_{on})(k_d + k_{off} + 1)}\left[k_{off}(k_{on} - k_1 k_d)\ln\frac{k_1}{k_{off}} + k_{off}(k_{on} - k_1 k_d)\ln\frac{k_{off}}{k_{on}}\right.$$
$$\left. + k_{off}(k_{on} - k_1 k_d)\ln k_d\right]$$

$$= \frac{k_{off}(k_{on} - k_1 k_d)\ln\frac{k_1 k_d}{k_{on}}}{(k_1 + k_{off} + k_{on})(k_d + k_{off} + 1)} \quad (S4)$$

$dS_{med}/dt|_{steady}$ first increases and then decreases with increasing $k_{off}$ peaking at,
$k_{off} = \sqrt{k_{on} k_d}$ (when $k_1 \rightarrow 0$).

## II. A scheme to calculate moments of the distribution, P(Q,t)

Consider the Master Eq.
$$\frac{dp_i(t)}{dt} = \sum_j L_{ij} p_j \qquad (S5)$$

L satisfies the condition, $\sum_i L_{ij} = 0$, which guarantees that $\sum_i p_i(t) = const$.

The joint distribution $\phi_i(Q,t)$, defined in the main text follows the kinetics (Eq. (4))

$$\frac{\partial \phi_i(Q,t)}{\partial t} = \sum_{j(\neq i)} \left[ w_i^j \phi_j(Q - \Delta s_i^j, t) - w_j^i \phi_i(Q,t) \right]$$

P(Q,t) is calculated by defining a moment generating function,

$$\psi_i(\lambda,t) = \int dQ \, e^{\lambda Q} \phi_i(Q,t) \qquad (S6)$$

which follows the a set of linear equations given by,

$$\frac{\partial \psi_i(\lambda,t)}{\partial t} = \sum_j H_{ij}(\lambda) \psi_j(\lambda,t) \qquad (S7)$$

where, $H_{ij}(\lambda) = (w_i^j)^{1-\lambda}(w_j^i)^{\lambda}(1-\delta_{ij}) - \delta_{ij} \sum_{j' \neq i} w_{j'}^i$.

Note, $H_{ij}(\lambda = 0) = L_{ij}$.

Eq. (S7) can become non-trivial to solve even when the Master equation (Eq. (S5)) can be solved analytically. However, it is possible to solve for the moments of Q(t) using the solutions of the Master equation and thus avoid the direct solution of Eq. (S7). The scheme is described below.

$$\left. \frac{\partial^n \psi_i(\lambda,t)}{\partial t^n} \right|_{\lambda=0} = \int dQ \, Q^n \phi_i(Q,t) = \langle Q^n \rangle_i$$

which gives the nth moment of the entropy exchanged by the state i until time t. Therefore,

$$\sum_i \left. \frac{\partial^n \psi_i(\lambda,t)}{\partial t^n} \right|_{\lambda=0} = \langle Q^n \rangle \text{ gives the nth moments of the total entropy exchanged up to time t.}$$

These moments can be calculated from Eq. (S7) recursively as follows.

*Calculation of $\langle Q \rangle$(t):*

Taking the derivative of Eq. (S7) with $\lambda$ produces,

$$\frac{\partial}{\partial t}\left(\frac{\partial \psi_i(\lambda,t)}{\partial \lambda}\right) = \sum_j H_{ij}(\lambda)\frac{\partial \psi_j(\lambda,t)}{\partial \lambda} + \sum_j \frac{\partial H_{ij}}{\partial \lambda}\psi_j(\lambda,t)$$

Setting $\lambda = 0$ on both the sides of the above equation we get,

$$\frac{\partial}{\partial t}\left(\frac{\partial \psi_i(\lambda,t)}{\partial \lambda}\right)\bigg|_{\lambda=0} = \sum_j H_{ij}(\lambda)\frac{\partial \psi_j(\lambda,t)}{\partial \lambda}\bigg|_{\lambda=0} + \sum_j \frac{\partial H_{ij}}{\partial \lambda}\psi_j(\lambda,t)\bigg|_{\lambda=0}$$

$$\Rightarrow \frac{\partial \langle Q \rangle_i}{\partial t} = \sum_j H_{ij}(\lambda=0)\langle Q \rangle_i + f_i(t) = \sum_j L_{ij}\langle Q \rangle_j + f_i(t) \qquad (S8)$$

where, $f_i(t) = \sum_j \frac{\partial H_{ij}}{\partial \lambda}\psi_j(\lambda,t)\bigg|_{\lambda=0} = \sum_j \frac{\partial H_{ij}}{\partial \lambda}\bigg|_{\lambda=0}\psi_j(\lambda=0,t) = \sum_j \frac{\partial H_{ij}}{\partial \lambda}\bigg|_{\lambda=0} p_j(t)$

In deriving the last equation I used $\psi_j(\lambda=0,t) = p_j(t)$, which follows from Eq. S6.
Therefore, one can calculate $f_i(t)$ from the solution of the Master equation in Eq. (S5).
Summing over all the states in Eq. (S8) we get,

$$\sum_i \frac{\partial \langle Q \rangle_i}{\partial t} = \sum_i \left[ L_{ij}\langle Q \rangle_j \right] + \sum_i f_i(t) = 0 + \sum_i f_i(t) = f_T(t)$$

$$\Rightarrow \frac{\partial \langle Q \rangle}{\partial t} = f_T(t) \qquad (S9)$$

This is subject to the initial condition, $\langle Q \rangle(t=0) = 0$.

$f_T$ can be calculation from the solutions of Eq. (S5).

*Calculation of $<Q^2>(t)$:*

Taking the second derivative of Eq. (S7) with $\lambda$ produces,

$$\frac{\partial}{\partial t}\left(\frac{\partial^2 \psi_i(\lambda,t)}{\partial \lambda^2}\right) = \sum_j H_{ij}(\lambda)\frac{\partial^2 \psi_j(\lambda,t)}{\partial \lambda^2} + \sum_j \frac{\partial^2 H_{ij}}{\partial \lambda^2}\psi_j(\lambda,t) + 2\sum_j \frac{\partial H_{ij}}{\partial \lambda}\frac{\partial \psi_j}{\partial \lambda}$$

Setting $\lambda = 0$ on both the sides of the above equation we get,

$$\frac{\partial \langle Q^2 \rangle_i}{\partial t} = \sum_j H_{ij}(\lambda=0)\langle Q^2 \rangle_i + \sum_j \frac{\partial^2 H_{ij}}{\partial \lambda^2}\bigg|_{\lambda=0}\psi_j(\lambda=0,t) + 2\sum_j \frac{\partial H_{ij}}{\partial \lambda}\bigg|_{\lambda=0}\frac{\partial \psi_j}{\partial \lambda}\bigg|_{\lambda=0}$$

$$\Rightarrow \frac{\partial \langle Q^2 \rangle_i}{\partial t} = \sum_j L_{ij}\langle Q^2 \rangle_i + f_i^{(2)}(t)$$

where,

$$f_i^{(2)}(t) = \sum_j \left.\frac{\partial^2 H_{ij}}{\partial \lambda^2}\right|_{\lambda=0} \psi_j(\lambda=0,t) + 2\sum_j \left.\frac{\partial H_{ij}}{\partial \lambda}\right|_{\lambda=0} \left.\frac{\partial \psi_j}{\partial \lambda}\right|_{\lambda=0} = \sum_j \left.\frac{\partial^2 H_{ij}}{\partial \lambda^2}\right|_{\lambda=0} p_j(t) + 2\sum_j \left.\frac{\partial H_{ij}}{\partial \lambda}\right|_{\lambda=0} \langle Q \rangle_j(t)$$

which can be calculated from the solutions of Eqs. (S5) and (S8).

Summing over all the states,

$$\frac{\partial \langle Q^2 \rangle}{\partial t} = f^{(2)}(t) \tag{S10}$$

where, $f^{(2)}(t) = \sum_i f_i^{(2)}(t)$ and the initial condition is, $\langle Q^2 \rangle_i (t=0) = 0$.

In the same way the higher order moments can be calculated recursively.

## III. Derivation of the Master equation with a fixed entropy exchange limit

I consider the three state minimal model described in the main text for this derivation. The resulting equations can be generalized for multiple states. The minimal model is described by the following first order reactions.

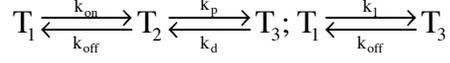

$$T_1 \underset{k_{off}}{\overset{k_{on}}{\rightleftarrows}} T_2 \underset{k_d}{\overset{k_p}{\rightleftarrows}} T_3;\ T_1 \underset{k_{off}}{\overset{k_1}{\rightleftarrows}} T_3$$

The kinetics is described by the Master Equation,

$$\frac{dp_i}{dt} = \sum_{j=1}^{3} L_{ij} p_j = \sum_{j(j \neq i)}^{3} (w_i^j p_j - w_j^i p_i)$$

$w_i^j$ denotes the transition rate of change state $j$ to state $i$. The w matrix is given by,

$$\begin{bmatrix} w_1^1 & w_1^2 & w_1^3 \\ w_2^1 & w_2^2 & w_2^3 \\ w_3^1 & w_3^2 & w_3^3 \end{bmatrix} = \begin{bmatrix} 0 & k_{off} & k_{off} \\ k_{on} & 0 & k_d \\ k_1 & k_p & 0 \end{bmatrix}$$

The amount of entropy flowing *into* the system from the reservoir when the transition $j \rightarrow i$ is executed following the Master equation is given by,

$$\Delta s_i^j = \ln\left(\frac{w_i^j}{w_j^i}\right)$$

,and, the $\Delta s$ matrix is given by,

$$\begin{bmatrix} \Delta s_1^1 & \Delta s_1^2 & \Delta s_1^3 \\ \Delta s_2^1 & \Delta s_2^2 & \Delta s_2^3 \\ \Delta s_3^1 & \Delta s_3^2 & \Delta s_3^3 \end{bmatrix} = \begin{bmatrix} n.d. & \ln\left(\frac{k_{off}}{k_{on}}\right) & \ln\left(\frac{k_{off}}{k_1}\right) \\ \ln\left(\frac{k_{on}}{k_{off}}\right) & n.d. & \ln\left(\frac{k_d}{k_p}\right) \\ \ln\left(\frac{k_1}{k_{off}}\right) & \ln\left(\frac{k_p}{k_d}\right) & n.d. \end{bmatrix} = \begin{bmatrix} n.d. & -\Delta_1 & \Delta_3 \\ \Delta_1 & n.d. & -\Delta_2 \\ -\Delta_3 & \Delta_2 & n.d. \end{bmatrix}$$

where, $\Delta_1 = \ln(k_{on}/k_{off})$, $\Delta_2 = \ln(k_p/k_d)$, and, $\Delta_3 = \ln(k_{off}/k_1)$, and, n.d. $\equiv$ not defined.

Next, I proceed to describe the time evolution of the joint probability distribution $\varphi_i(Q,t|i_0,0,0)$, which is the conditional probability of the system being at the state $i$ at time $t$ after receiving a total amount of entropy Q from the reservoir in the time interval 0 to $t$ starting from an initial state $i_0$ with zero entropy received at time $t=0$. For brevity, I will denote $\varphi_i(Q,t|i_0,0,0)$ by $\varphi_i(Q,t)$ for rest of the calculation. The probability distribution, $\varphi_i(Q,t)$, follows the Master equation,

$$\frac{\partial \phi_1(Q,t)}{\partial t} = w_1^2 \phi_2(Q+\Delta_1,t) + w_1^3 \phi_3(Q-\Delta_3,t) - (w_2^1 + w_3^1)\phi_1(Q,t)$$

$$\frac{\partial \phi_2(Q,t)}{\partial t} = w_2^1 \phi_1(Q-\Delta_1,t) + w_2^3 \phi_3(Q+\Delta_2,t) - (w_1^2 + w_3^2)\phi_2(Q,t)$$

$$\frac{\partial \phi_3(Q,t)}{\partial t} = w_3^1 \phi_1(Q+\Delta_3,t) + w_3^2 \phi_2(Q-\Delta_2,t) - (w_1^3 + w_2^3)\phi_3(Q,t) \tag{S11}$$

The limit on the total amount of entropy ($E$) that can be exchanged with the reservoir is implemented by imposing a reflecting boundary condition in entropy exchange in the above equations. Thus, when a stochastic trajectory reaches the limit $E$, a reaction that releases entropy to the reservoir is executed, and, the stochastic trajectory is not lost forever as in the case of an absorbing boundary condition. The result of imposing the reflective boundary condition is analyzed for a simple example where specific rates are chosen such that the entropy exchanges can be described by changes on a regular lattice. Without any loss of generality a set of rates are chosen such that $\Delta_1>0$, $\Delta_2>0$, and, $\Delta_3>0$, and, and $\Delta_2/\Delta_1=p$ (integer) and $\Delta_3/\Delta_1=q$ (integer), and, the total entropy exchange at any time resides on a site (say $n$) on this lattice, $Q=Q_n=n\Delta_1$. I also assume, $\Delta_3>\Delta_2>\Delta_1$. Eq.(S11) now can be described on a grid of a unit entropy exchange $\Delta_1$ as,

$$\frac{\partial \phi_1(n,t)}{\partial t} = w_1^2 \phi_2(n+1,t) + w_1^3 \phi_3(n-q,t) - (w_2^1 + w_3^1)\phi_1(n,t)$$

$$\frac{\partial \phi_2(n,t)}{\partial t} = w_2^1 \phi_1(n-1,t) + w_2^3 \phi_3(n+p,t) - (w_1^2 + w_3^2)\phi_2(n,t)$$

$$\frac{\partial \phi_3(n,t)}{\partial t} = w_3^1 \phi_1(n+q,t) + w_3^2 \phi_2(n-p,t) - (w_1^3 + w_2^3)\phi_3(n,t) \tag{S12}$$

Eq. (S12) describes the time evolution of the states (1,2, or 3) as the system moves on the lattice (unit lattice size = $\Delta_1$) of the total entropy exchange. The step sizes for increasing (or decreasing) the total entropy exchange depends on the particular state undergoing the transition. The state 3 increases (or decreases) entropy exchange with a step size of $q$ (or $p$), the state 2 increases (or decreases) moves on the lattice with a step size of $p$ (or 1), and, the state 1 increases (or decreases) moves on the lattice with a step size of 1 (or $q$). Therefore, $\phi_1(n,t)$, $\phi_2(n,t)$ and $\phi_3(n,t)$ evolves on the lattice according to those rules.

A reflecting boundary condition at $n=E$ is imposed, for simplicity $E$ is taken to be a multiple of the least common multiple (lcm) of $p$ and $q$, so that all the states are able to access the limit exactly. The reflecting boundary condition demands(1),
$\phi_1(E+1,t) = \phi_1(E+2,t) = ..0\,; \phi_2(E+1,t) = \phi_2(E+2,t) = ..0\,; \phi_3(E+1,t) = \phi_3(E+2,t) = ..0\,;$

The time evolution at $n=E$ is given by,
$$\frac{\partial \phi_1(E,t)}{\partial t} = w_1^3 \phi_3(E-q,t) - (c_1 w_2^1 + c_2 w_3^1)\phi_1(E,t)$$
$$\frac{\partial \phi_2(E,t)}{\partial t} = w_2^1 \phi_1(E-1,t) - (c_3 w_1^2 + c_4 w_3^2)\phi_2(E,t)$$
$$\frac{\partial \phi_3(E,t)}{\partial t} = w_3^2 \phi_2(E-p,t) - (c_5 w_1^3 + c_6 w_2^3)\phi_3(E,t)$$

(S13)

The the parameters, $c_1,..,c_6$, in Eq. (S13) are introduced to determine the transition rates at which the system leaves once it reaches the limit at E. The parameters, $c_1,..,c_6$, are determined by using the condition that the total probability is conserved in the time evolution. This condition holds for reflecting boundary conditions where no stochastic trajectory is lost. We define variables,
$$\bar{\phi}_1(t) = \sum_{n=-\infty}^{\infty} \phi_1(n,t)\,, \bar{\phi}_2(t) = \sum_{n=-\infty}^{\infty} \phi_2(n,t)\,, \bar{\phi}_3(t) = \sum_{n=-\infty}^{\infty} \phi_3(n,t)$$
and the above condition implies,
$$\frac{\partial(\bar{\phi}_1(t)+\bar{\phi}_2(t)+\bar{\phi}_3(t))}{\partial t} = 0$$
From Eqs. (S12) and (S13),
$$\frac{\partial \bar{\phi}_1(t)}{\partial t} = \sum_{n=-\infty}^{E} \left( w_1^2 \phi_2(n+1,t) + w_1^3 \phi_3(n-q,t) - (w_2^1 + w_3^1)\phi_1(n,t) \right)$$
$$= \sum_{n=-\infty}^{E-1} w_1^2 \phi_2(n+1,t) + \sum_{n=-\infty}^{E} w_1^3 \phi_3(n-q,t) - \sum_{n=-\infty}^{E-1} (w_2^1 + w_3^1)\phi_1(n,t) - (c_1 w_2^1 + c_2 w_3^1)\phi_1(E,t)$$
$$= w_1^2 \bar{\phi}_2(t) + w_1^3 \bar{\phi}_3(t) - w_1^3 \phi_3(E,t) - (w_{21} + w_{31})\bar{\phi}_1(t) - ((c_1-1)w_2^1 + (c_2-1)w_3^1)\phi_1(E,t)$$
In deriving the last step I have used the fact that state 3 increases entropy exchange with a step size of $q$. Similarly, we can now derive the equations for $\phi_2$ and $\phi_3$.
$$\frac{\partial \bar{\phi}_2(t)}{\partial t} = \sum_{n=-\infty}^{E} \left( w_2^1 \phi_1(n-1,t) + w_2^3 \phi_3(n+p,t) - (w_1^2 + w_3^2)\phi_2(n,t) \right)$$
$$= w_2^1 \bar{\phi}_1(t) - w_2^1 \phi_1(E,t) + w_2^3 \bar{\phi}_3(t) - (w_1^2 + w_3^2)\bar{\phi}_2(t) - \left((c_3-1)w_1^2 + (c_4-1)w_3^2\right)\phi_2(E,t)$$

$$\frac{\partial \bar{\phi}_3(t)}{\partial t} = \sum_{n=-\infty}^{E} \left( w_3^1 \phi_1(n+q,t) + w_3^2 \phi_2(n-p,t) - (w_1^3 + w_2^3)\phi_3(n,t) \right)$$

$$= w_3^1 \bar{\phi}_1(t) + w_3^2 \bar{\phi}_2(t) - w_3^2 \phi_2(E,t) - (w_1^3 + w_2^3)\bar{\phi}_3(t) - ((c_5-1)w_1^3 + (c_6-1)w_2^3)\phi_3(E,t)$$

Therefore,

$$\frac{\partial \left[\bar{\phi}_1(t) + \bar{\phi}_2(t) + \bar{\phi}_3(t)\right]}{\partial t} = -w_1^3 \phi_3(E,t) - ((c_1-1)w_2^1 + (c_2-1)w_3^1)\phi_1(E,t)$$

$$- w_2^1 \phi_1(E,t) - \left((c_3-1)w_1^2 + (c_4-1)w_3^2\right)\phi_2(E,t)$$

$$- w_3^2 \phi_2(E,t) - ((c_5-1)w_1^3 + (c_6-1)w_2^3)\phi_3(E,t)$$

If the right hand side of the above equation is set zero as required by the conservation of the total probability, then the parameters assume the values,

$c_1=0$, $c_2=1$, $c_3=1$, $c_4=0$, $c_5=0$, $c_6=1$. This shows that at $n=E$, any reaction step that requires flow of entropy from the reservoir are not executed (or the transition probabilities are zero). Thus the kinetics at $n=E$ is given by,

$$\frac{\partial \phi_1(E,t)}{\partial t} = w_1^3 \phi_3(E-q,t) - w_3^1 \phi_1(E,t)$$

$$\frac{\partial \phi_2(E,t)}{\partial t} = w_2^1 \phi_1(E-1,t) - w_1^2 \phi_2(E,t)$$

$$\frac{\partial \phi_3(E,t)}{\partial t} = w_3^2 \phi_2(E-p,t) - w_2^3 \phi_3(E,t)$$

(S14)

## IV. Exact solution for the joint distribution $\phi_i(Q,t)$ for a simple case

I consider the system with one TCR and one pMHC here as described in section I. The transition, $T_3 \to T_1$ (the KPR step) is executed at a rate, $k'_{off}$. The rate constants are chosen as, $k_{on}=\exp(-1)=1/e$, $k_{off}=\exp(-2)=1/e^2$, $k_p=1$, $k_d=\exp(-1)=1/e$, and, $k_1=\exp(-2)=1/e^2$ and $k'_{off}=\exp(-1)=1/e$. The above choice of the rates makes the entropy exchanges in each reactions integer valued, thus, as the system evolves in time, the medium entropy exchange moves the system on a lattice of size 1. Two reflective boundary conditions at $Q=E=3$ and $Q=E=1$ are imposed. This keeps the system confined within 9 states, each state denoting the pair $(i,Q)$, where, the chemical state of the complex ($T_1$, $T_2$, or $T_3$) is designated by $i$ and the entropy exchanged is given by $Q$ (1, 2, or, 3). In this case, with a finite number of states in the kinetics, Eq.(4) is amenable to analytical methods. Eq. (S12) and Eq. (S14) calculated at the reflective boundaries of $E=3$ and $E=1$ can be used to describe the above kinetics which is summarized by the linear equation

$$\frac{\partial \varphi_\alpha}{\partial t} = \sum_\beta S_{\alpha\beta} \varphi_\alpha \qquad (S15)$$

In the above equation, each $\alpha$ corresponds to a state, $(i,Q)$, and $\alpha$ assumes integer values 1 to 9. The 9x9 matrix $S$ contains elements given by Eq. (S12) and Eq. S(14) corresponding the two boundary conditions. Eq. (S15) can be solved by standard methods that involve calculation of the eigenvector and eigenvalues of $S$. I briefly describe the method below.

$S_{\alpha\beta} = \langle \alpha | L | \beta \rangle$. If the right and left eigenvectors of $S$ are, $|R_n\rangle$ and $\langle L_n|$, respectively, such that, $S|R_n\rangle = \lambda_n |R_n\rangle$ and $\langle L_n|S = \lambda_n \langle L_n|$, then we can expand $\varphi_\alpha = \langle \alpha | \varphi \rangle$ as,

$|\varphi(t)\rangle = \sum_n a_n(t) |R_n\rangle$. Thus, we can write Eq. (S15) as,

$$\frac{\partial \langle \alpha | \varphi \rangle}{\partial t} = \sum_\beta \langle \alpha | S | \beta \rangle \langle \beta | \varphi \rangle$$

$$\Rightarrow \frac{\partial \langle \alpha | R_n \rangle b_n^{-1} \langle L_n | \varphi \rangle}{\partial t} = \sum_\beta \langle \alpha | R_n \rangle b_n^{-1} \langle L_n | L | R_m \rangle b_m^{-1} \langle L_m | \beta \rangle \langle \beta | \varphi \rangle$$

where, repeated n and m indices are summed over, and we use the identity

$$\sum_n |R_n\rangle b_n^{-1} \langle L_n| = 1 \text{ ,or, } \langle L_n | R_n \rangle b_n^{-1} \langle L_n | R_n \rangle = \langle L_n | R_n \rangle \Rightarrow \langle L_n | R_n \rangle = b_n$$

The vectors are orthogonal to each other, i.e.,
$\langle L_n | R_m \rangle = b_n \delta_{mn}$

Since $|R_n\rangle$ and $\langle L_n|$ are the eigenvectors, we get from the above equation,

$$\frac{\partial \langle\alpha|R_n\rangle b_n^{-1}\langle L_n|\varphi\rangle}{\partial t} = \sum_\beta \langle\alpha|R_n\rangle b_n^{-1}\lambda_n\langle L_n|\beta\rangle\langle\beta|\varphi\rangle$$

$$\Rightarrow \frac{\partial \langle\alpha|R_n\rangle b_n^{-1}\langle L_n|\varphi\rangle}{\partial t} = \langle\alpha|R_n\rangle b_n^{-1}\lambda_n\langle L_n|\varphi\rangle$$

$$\Rightarrow \frac{\partial \langle\alpha|R_n\rangle b_n^{-1} a_n(t)}{\partial t} = \langle\alpha|R_n\rangle b_n^{-1}\lambda_n a_n(t)$$

Since the eigenvectors form a complete orthogonal set, the equality will be valid for each term in the above equation, i.e.,

$$\frac{\partial a_n(t)}{\partial t} = \lambda_n a_n(t) \qquad (S16)$$

As, $\langle\alpha|\varphi(0)\rangle = \varphi_\alpha(0)$ is given by the initial condition that the system starts from the state $T_1$ at $Q=1$ at $t=0$. This is used to calculate $a_n(0)$ from $|\varphi(0)\rangle = \sum_n a_n(0)|R_n\rangle$

The solution for Eq. (S16) is given by, $a_n(t) = a_n(0)e^{\lambda_n t}$

Thus, $\varphi_\alpha(t) = \langle\alpha|\varphi(t)\rangle = \sum_n \langle\alpha|R_n\rangle b_n^{-1}\langle L_n|\varphi(t)\rangle = \sum_n \langle\alpha|R_n\rangle b_n^{-1} a_n(0)e^{\lambda_n t}$ (S17)

Using the above solution the probability of the particle to be in a particular chemical state is calculated as,

$$p_i(t) = \sum_{Q=1}^{3} \phi_i(Q,t) .$$

The above equations can be solved using Mathematica (code available at http://planetx.nationwidechildrens.org/~jayajit). The solution for the above example is given by,

$$p_2(t) = (1 - e^{-yt})/(ye) \qquad (S17a)$$

$$p_3(t) = (1 + e^2 + e^{1-yt}(1 - e^{t+t/e^2+1}y))/(y(1+e^2)) \qquad (S17b)$$

and, $p_1(t) = 1 - p_2(t) - p_3(t)$
where, $y = (1 + e + e^2)/e$ .

The comparisons of Eqs. (S17a,b) with the corresponding MC simulations are shown in Fig. 2B.

**Table S1: Details of the parameter values used in the simulations**

The values obtained from the literature were converted to the units of time and length scales in seconds and microns, respectively.

| Parameter | Value | Reference | Comments |
|---|---|---|---|
| $k_{on}$ | $0.003 s^{-1}$ | (2) | The measured 3D $k_{on}$ rate for a 2B4 TCR binding to MCC peptide-MHC is converted to a 2D binding rate by using $(k_{on})_{2D}=(k_{on})_{3D}/d$, where, $d$ (=1.2nm) is the length scale where the interaction takes place(3). |
| $k_{off}$ | varied from $0.001s^{-1}$ to $10s^{-1}$ | (4) | Peptide fragments derived from pathogens eliciting a strong T cell activation have larger half-lives (~20s or larger), whereas, self-peptides bind fleetingly (half-lives 0.1s or smaller) to the TCR. |
| $k_1$ | $10^{-8} s^{-1}$ | NA | A small value was chosen to keep the entropy calculations finite. This reaction rarely occurred in the time scales relevant for T cell activation (<30mins). |
| $k_p$ | $1.0 s^{-1}$ | (5) | The tyrosine residues in the ITAMs are phosphorylated by Src kinases with a catalytic rate $5s^{-1}$. |
| $k_d$ | $0.1 s^{-1}$ | (5) | Dephosphorylation of activated ITAMs is carried by the phosphatase SHP1. The enzymatic reaction is approximated by a first order reaction (rate = $k_{cat}$[SHP1]/$K_M$) Using the measured binding (330,000 $M^{-1}.s^{-1}$), unbinding (0.05s-1), and the catalytic rate ($k_{cat}$=5s-1) for SHP1 acting on a substrate (Lck) and assuming [SHP1]=1µM, one gets $k_d$~0.3 $s^{-1}$. Measurements of dephosphorylation rates for activated ITAMs associated with Fc receptors give a first order rate ~ $0.1s^{-1}$-$0.05s^{-1}$(6). |
| $N_{T0}$ | 100 molecules/µm$^2$ | (3) | Ref. (3) estimates 50,000 TCRs to be present in a 2B4 T cell of surface area 500µm$^2$. |
| $N_{M0}$ | 100 molecules/µm$^2$ | (7) | ~$10^5$ – $10^6$ MHC molecules are present in a single cell. Numbers vary depending on the cell type. Assuming a radius of 10µm for the APCs this gives a range of MHC concentration 80-800 molecules/ µm$^2$ per cell. A fraction of these MHCs will be occupied by peptides. |

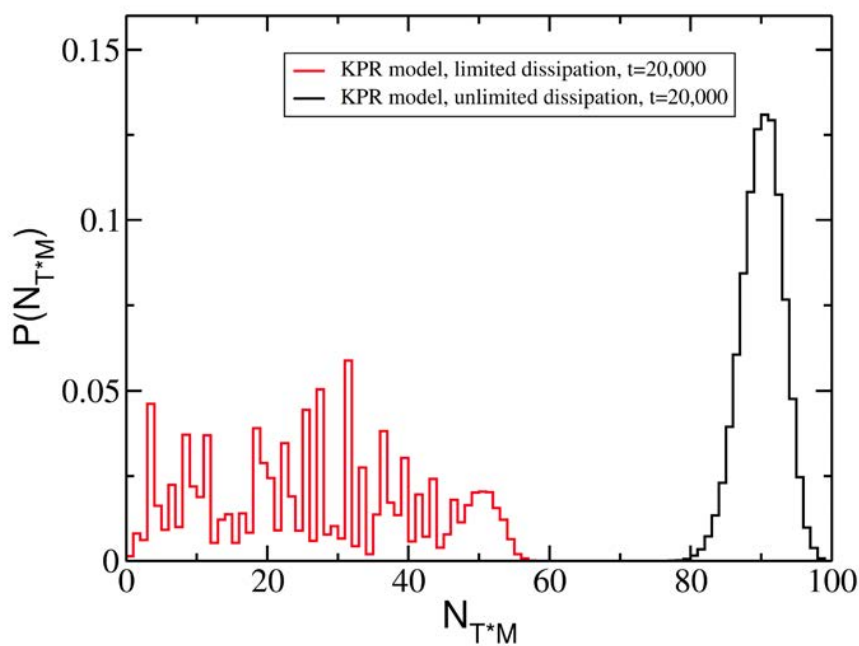

**Fig. S1.** Distributions of $N_{T*M}$ at t=20,000s for the dissipation limited (E=500) and the unlimited dissipation cases. $P(N_{T*M})$ are calculated using $10^6$ stochastic trajectories. In the simulations, $k_{off}$=0.001s$^{-1}$, and, rest of the parameters are given in the Materials and Methods section.

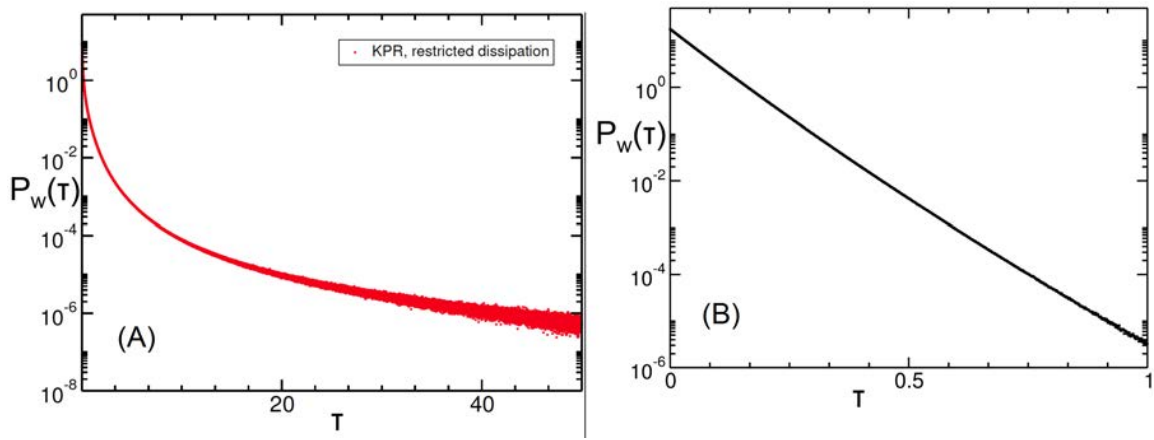

**Fig. S2.** Waiting time distributions, $P_w(\tau)$, for the limited (A) and the unlimited dissipation (B) cases, calculated using over $10^4$ stochastic trajectories. The parameters are the same as in Fig. S1. The data in (A) show a non-Debye decay but cannot be fitted well with a stretched exponential form. In contrast, the data in (B) show an exponential decay.

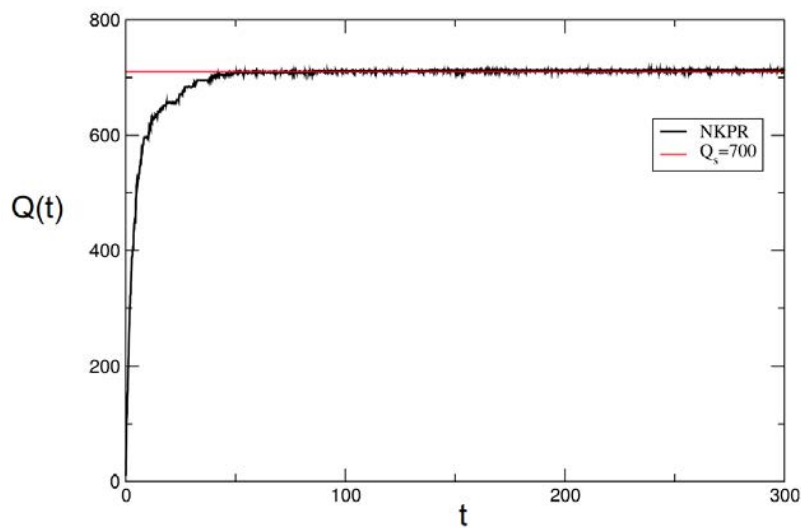

**Fig. S3**. Kinetics of the total medium entropy received by the system (Q(t)), corresponding to a single stochastic trajectory, for the NKPR model. Q(t) reaches saturation (~$Q_s$) after a short time scale.

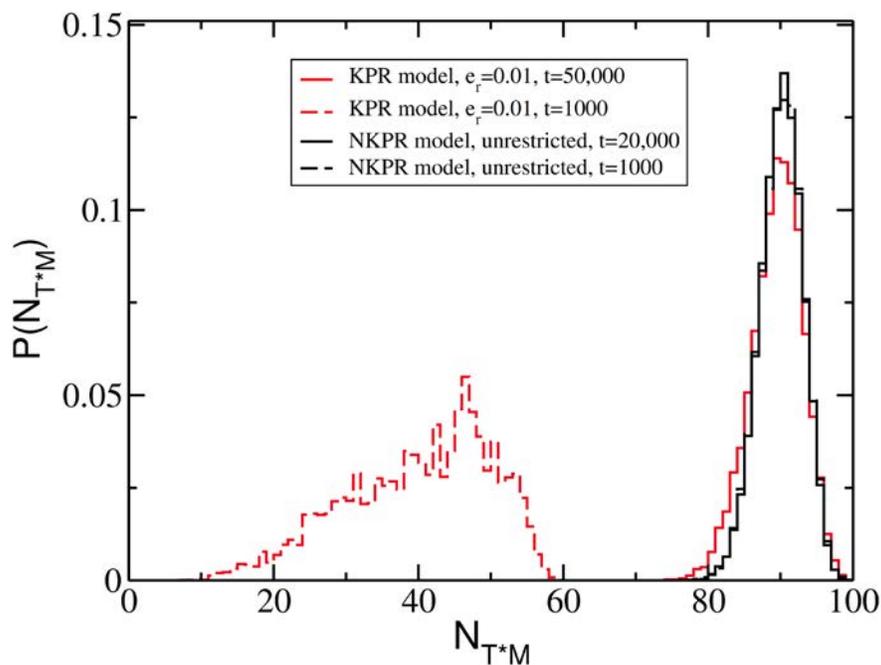

**Fig. S4.** Distributions of $N_{T*M}$ KPR model at $e_r=0.01s^{-1}$. The distributions are calculated at times $t<\tau_{trans}$ (dashed red line) and $t>\tau_{trans}$ (solid red line) using over $10^4$ stochastic trajectories. The data for the NKPR model are shown for comparison. In the simulations, $k_{off}=0.001s^{-1}$, and, rest of the parameters are given in the Materials and Methods section.

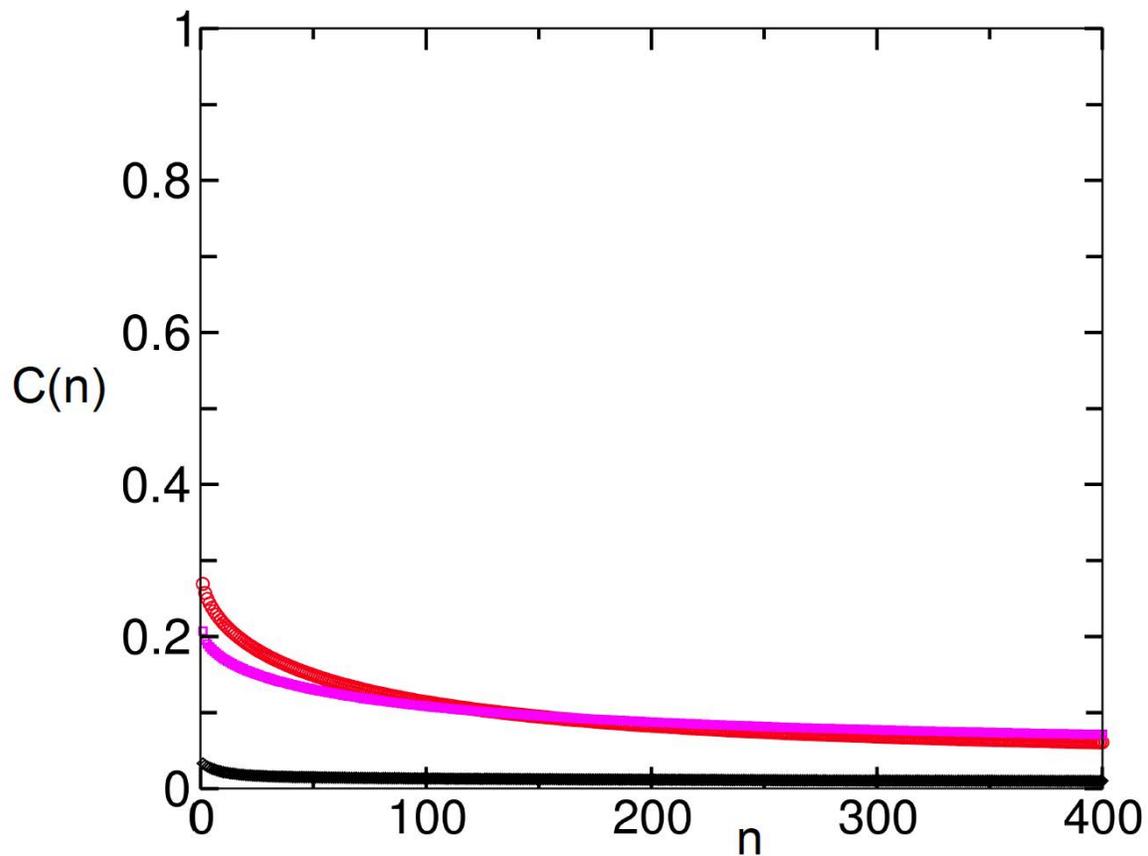

**Fig. S5.** C(n) vs n, calculated using over $10^4$ stochastic trajectories, for the KPR model for $e_r=0.01s^{-1}$ for times $t<\tau_{trans}$ (red circle) and $t>\tau_{trans}$ (magenta squares). The data for the NKPR model (black diamonds) are shown for comparison.

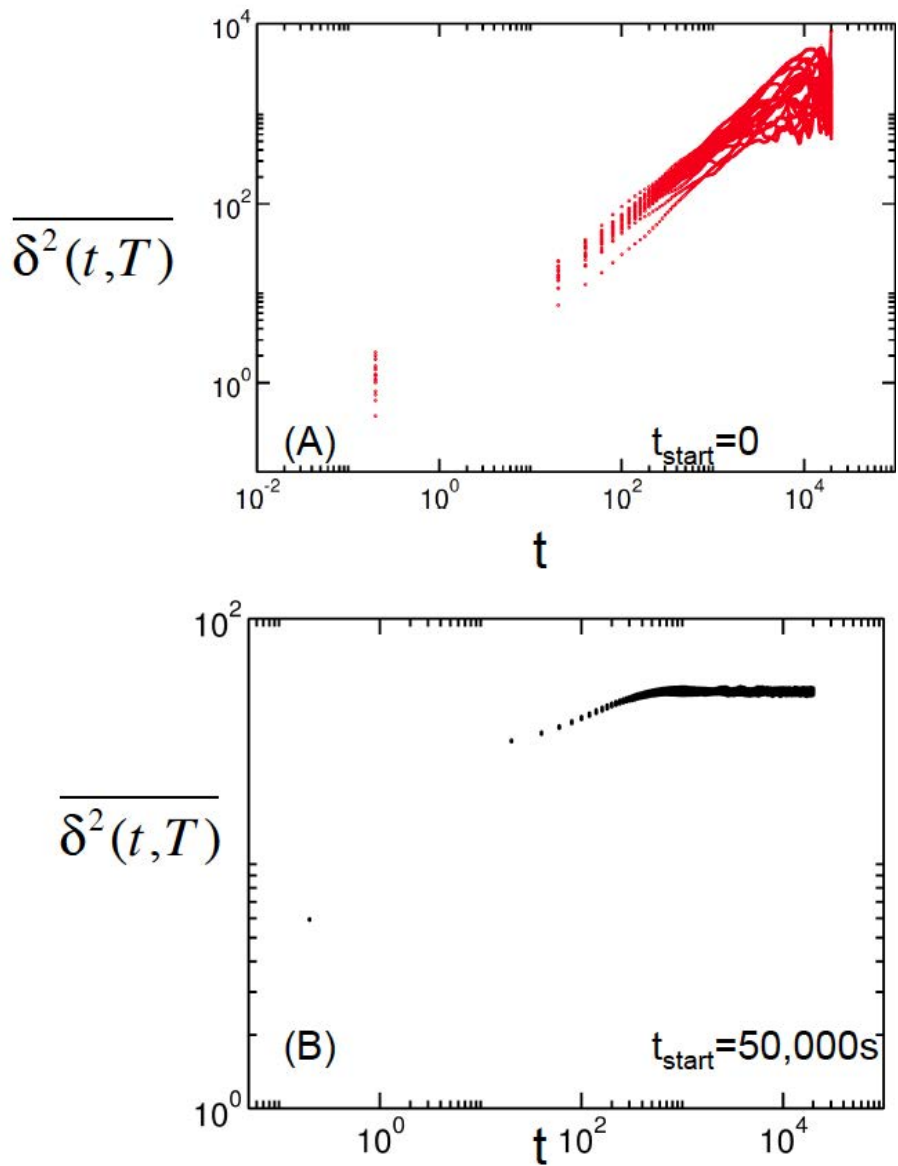

**Fig. S6.** Variation of $\overline{\delta^2(t,T)}$ with t for 20 different stochastic trajectories for the KPR model for $e_r=0.01s^{-1}$ for times $t<\tau_{trans}$ (A) and $t>\tau_{trans}$ (B). (A) $t_{start}=0$ and data are collected until t=20,000. (B) $t_{start}=50,000s$, and, the data are collected until $t=10^6 s$.

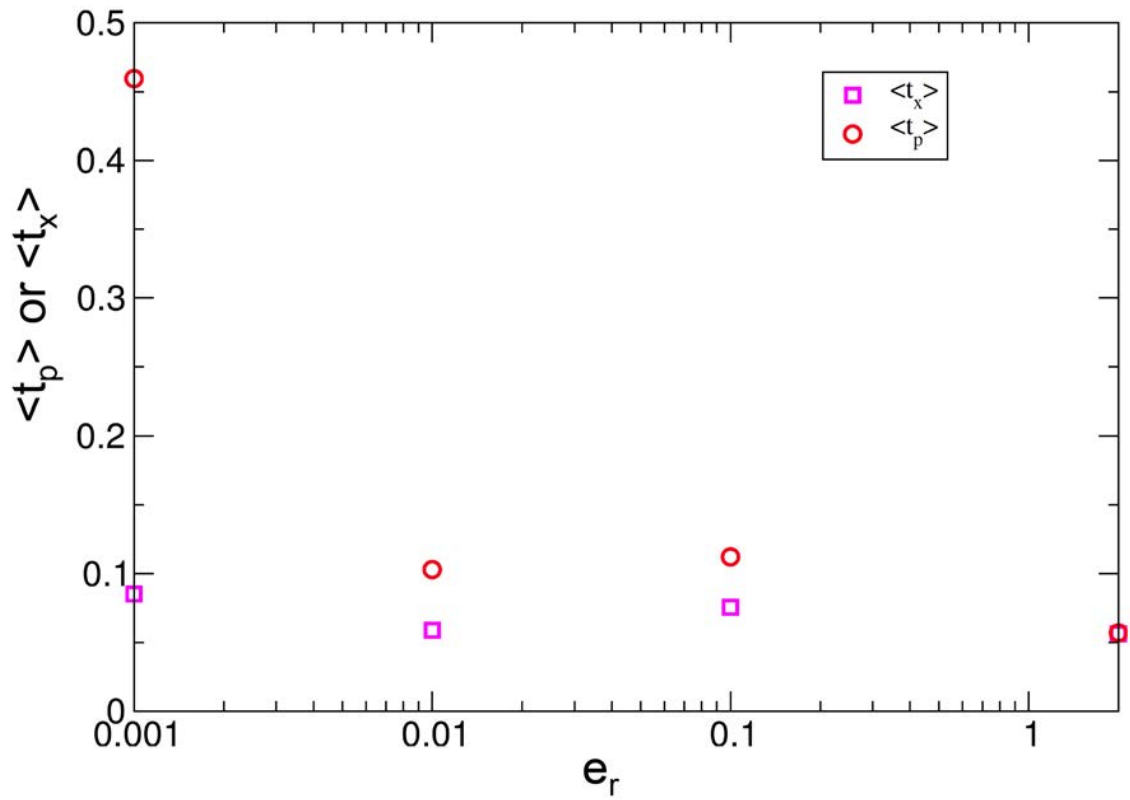

**Fig. S7.** Variation of $\langle t_p \rangle$ and $\langle t_x \rangle$ with $e_r$ for the KPR model. $P(t_p)$ and $P(t_x)$ for each $e_r$ is calculated for times $t < \tau_{trans}$ using over $10^4$ stochastic trajectories.

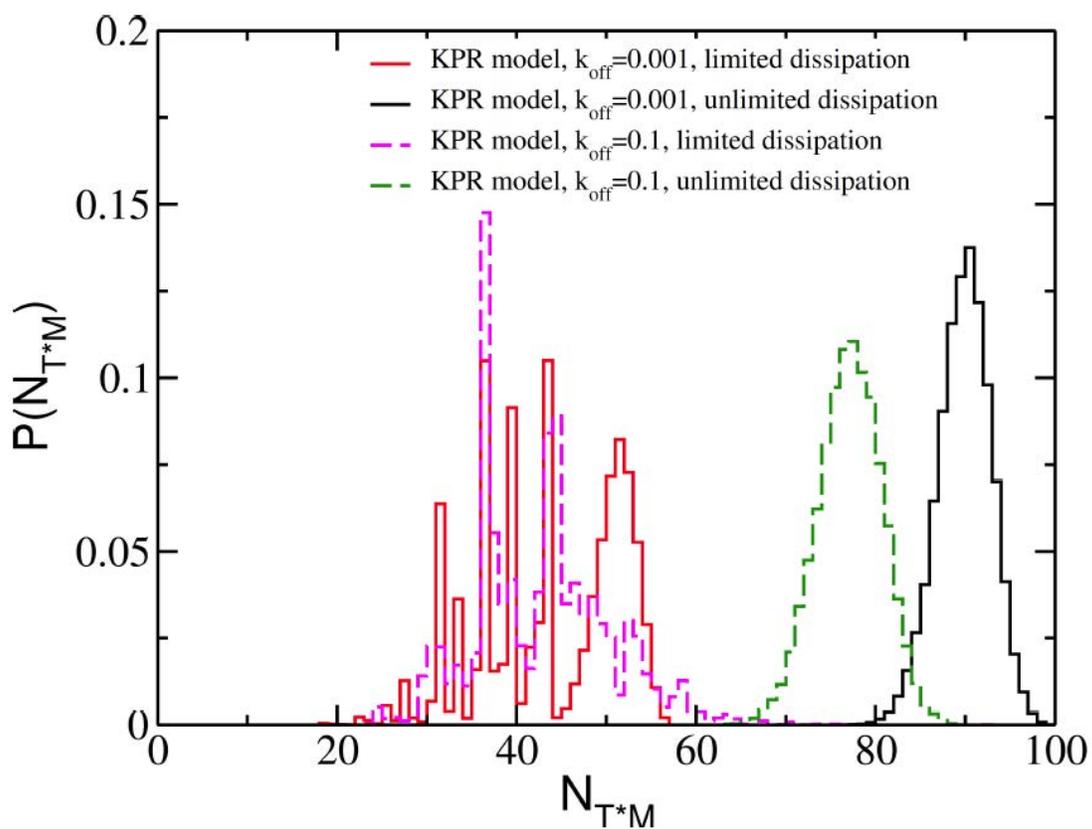

**Fig. S8.** Distributions of $N_{T*M}$ for the KPR model calculated using over $10^4$ stochastic trajectories at t=300s. Data are shown for cases at a fixed dissipation limit (E=500) for $k_{off}$=0.001s$^{-1}$ and $k_{off}$=0.1s$^{-1}$. The data for the unlimited dissipation cases for the same ligand affinities are shown for comparison. In the presence of limited dissipation $P(N_{T*M})$ for different ligand affinities show substantial overlap.

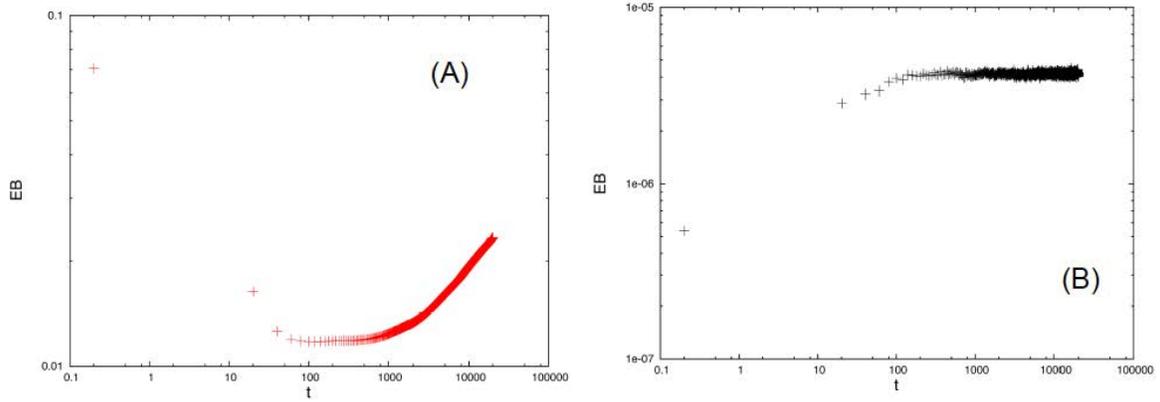

**Fig. S9. Variation of the entropy breaking parameter EB with the time interval t.**
EB is defined as, EB=$\lim_{T\to\infty}$ ⟨$\xi^2$⟩ -⟨$\xi$⟩$^2$, where, $\xi = \overline{\delta^2(t,T)}/\langle\overline{\delta^2(t,T)}\rangle$. Thus,

$$EB = \lim_{T\to\infty} \frac{\langle(\overline{\delta^2(t,T)})^2\rangle}{\langle\overline{\delta^2(t,T)}\rangle^2} - 1$$

, and EB was calculated from the ensemble of $\overline{\delta^2(t,T)}$ generated from single stochastic trajectories in the simulation. (A) EB shown for the case with a fixed dissipation limit at E=500 calculated using over 500 single trajectories. The parameters in the simulations are the same as that of Fig. 2D. (B) EB shown for the case with no limit on dissipation, the simulation parameters are the same as in Fig. 2D. Note the small values of EB in this case compared to (B).

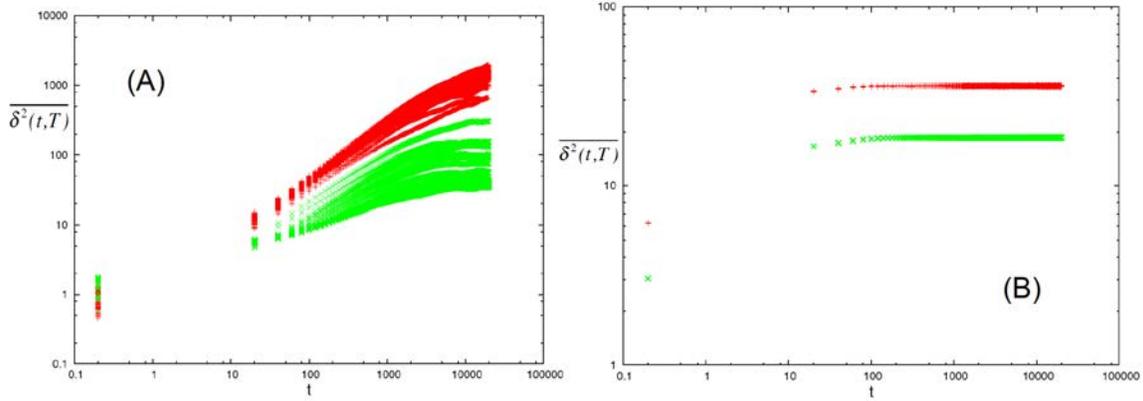

**Fig. S10. Effect of the total number of receptors and ligands on the time averaged MSD.** The apparent saturation of $\overline{\delta^2}(T,t)$ depends on the upper bound on the position of the random walker in the CTRW which are determined by the largest values of $N_{TM}$ and $N_{T*M}$ in the simulations. The largest values of $N_{TM}$ and $N_{T*M}$ in turn are determined by the total numbers of T (or $N_{T0}$) and M molecules (or $N_{M0}$) in the system by the conservation laws, $N_{T0} = N_T + N_{TM} + N_{T*M}$ and $N_{M0} = N_M + N_{TM} + N_{T*M}$. (A) Variation of $\overline{\delta^2}(T,t)$ with t for the dissipation limited case for $N_{T0}=N_{M0}=$ 100 (red) and $N_{T0}=N_{M0}=$ 50(green). The other parameters are the same as in Fig. 2D. (B) Variation of $\overline{\delta^2}(T,t)$ with t for the kinetics without any limit on dissipation for $N_{T0}=N_{M0}=$ 100 (red) and $N_{T0}=N_{M0}=$ 50(green). The other parameters are the same as in Fig. 2D. 100 different trajectories for each case are shown in both the figures.